\documentclass[lettersize,journal]{IEEEtran}
\usepackage[utf8]{inputenc}
\usepackage{amsmath,amsfonts,amssymb,amsthm}
\usepackage{algorithmic}
\usepackage{algorithm}
\usepackage{array}
\usepackage{bm}
\usepackage{mathtools}
\usepackage{multirow}
\usepackage[mathscr]{eucal}
\usepackage{threeparttable}
\usepackage{makecell}
\usepackage{enumitem}
\usepackage[table]{xcolor}
\usepackage{booktabs}
\usepackage[caption=false,font=normalsize,labelfont=sf,textfont=sf]{subfig}
\usepackage{textcomp}
\usepackage{stfloats}
\usepackage{url}
\usepackage{verbatim}
\usepackage{graphicx}
\usepackage{cite}
\usepackage{soul}
\usepackage[hidelinks]{hyperref}

\newcommand{\black}[1]{\textcolor{black}{#1}}

\newcommand{\yu}[1]{\black{#1}}
\newcommand{\zeng}[1]{\black{#1}}
\newcommand{\ie}{i.e.}
\newcommand{\eg}{e.g.}

\def\figureautorefname~#1\null{Fig.~#1\null}
\def\equationautorefname~#1\null{Eq.~(#1)\null}
\def\sectionautorefname~#1\null{Section~#1\null}
\def\subsectionautorefname~#1\null{Section~#1\null}
\def\subsubsectionautorefname~#1\null{Section~#1\null}
\def\algorithmautorefname~#1\null{Algorithm~#1\null}

\allowdisplaybreaks[4]

\newtheorem{theorem}{Theorem}

\begin{document}

\title{Towards the Readability of LLM-Generated Codes through Multitask Representation Engineering}

\author{Huifan Gao,
        Liuhua He,
        Yinghui Pan,
        Shenbao Yu,
        Yifeng Zeng,
        Shengchao Qin,
        Weidi Sun
%\thanks{{Manuscript received ?/?/?; revised ?/?/?; accepted ?/?/?. Date of publication ?/?/?; date of current version ?/?/?. This article was recommended by ?. (Corresponding author: Yifeng Zeng.)}}

\thanks{Huifan Gao is with the School of Aerospace Engineering, Xiamen University, Xiamen, China (e-mail:huifangao@foxmail.com).}

\thanks{Liuhua He is with the School of Artificial Intelligence, Shenzhen University, Shenzhen, China (e-mail:jcxghlhx@163.com).}

\thanks{Yinghui Pan is with the School of Artificial Intelligence, Shenzhen University, Shenzhen, China (e-mail: panyinghui@szu.edu.cn).}

\thanks{Shenbao Yu is with the College of Computer and Cyber Security, Fujian Normal University, Fuzhou, China (e-mail:yushenbao@fjnu.edu.cn).}

\thanks{Yifeng Zeng is with the Department of Computer \& Information Sciences, Northumbria University, UK (e-mail:yifeng.zeng@northumbria.ac.uk).}

\thanks{Shengchao Qin is with the School of Computer Science and Technology, Xidian University, Xi’an, China (e-mail:shengchao.qin@gmail.com).}

\thanks{Weidi Sun is with Peking University, Beijing, China (e-mail:weidisun@pku.edu.cn).}

}

% The paper headers. Replace the placeholder journal information before final submission.
\markboth{IEEE Journal Submission,~Vol.~XX, No.~XX, Month~2026}%
{Anonymous submission: Towards the Readability of LLM-Generated Codes Through Multitask Representation Engineering}

\maketitle

\begin{abstract}
Correctness and readability are key measures of code quality, respectively ensuring functional fidelity and ease of comprehension. While most existing research focuses on improving the correctness of large language models~(LLMs) generated codes, readability remains under-addressed. Enhancing readability through targeted control is challenging due to its subjective nature. 
In this article, we employ representation engineering~(RepE) as the targeted control method given its characteristics of low data dependency and low computational cost. Prior work on RepE has primarily focused on the targeted control for a single task, but improving the code readability requires the control across multiple tasks. Accordingly we proposes the multitask RepE framework  and theoretically discuss the impact of the multitask steering method on the tradeoff between the code readability and correctness. We further provide comprehensive experiments in support. All the relevant implementations are open-source and available upon request.
\end{abstract}


\section{Introduction}
\IEEEPARstart{I}{n} software engineering, code correctness and readability are both critical metrics for assessing code quality~\cite{spinellis2006code}. The code correctness refers to whether the code executes strictly according to predefined requirements and produces accurate results. The code readability, which is however subjective, refers to whether the code is easy to understand and maintain. What are comprehensive factors impacting the code readability is still an open question in the code quality research~\cite{fakhoury2019improving}. In this article, we mainly focus on the three commonly-recognized evaluation metric including ($a$) {\it comment density}, ($b$) {\it naming conventions}, and ($c$) {\it cyclomatic complexity}~\cite{buse2009learning,posnett2011simpler,scalabrino2017automatically}.These three metrics characterize code readability from complementary perspectives: comment density reflects the sufficiency of explanatory information, naming conventions affect whether identifiers convey clear semantics, and cyclomatic complexity measures the structural difficulty of understanding the control flow. Although they cannot cover all readability-related factors, they provide representative and measurable dimensions for analyzing the readability of LLM-generated code.
\begin{figure*}[!t]
    \centering
    \includegraphics[width=1\linewidth]{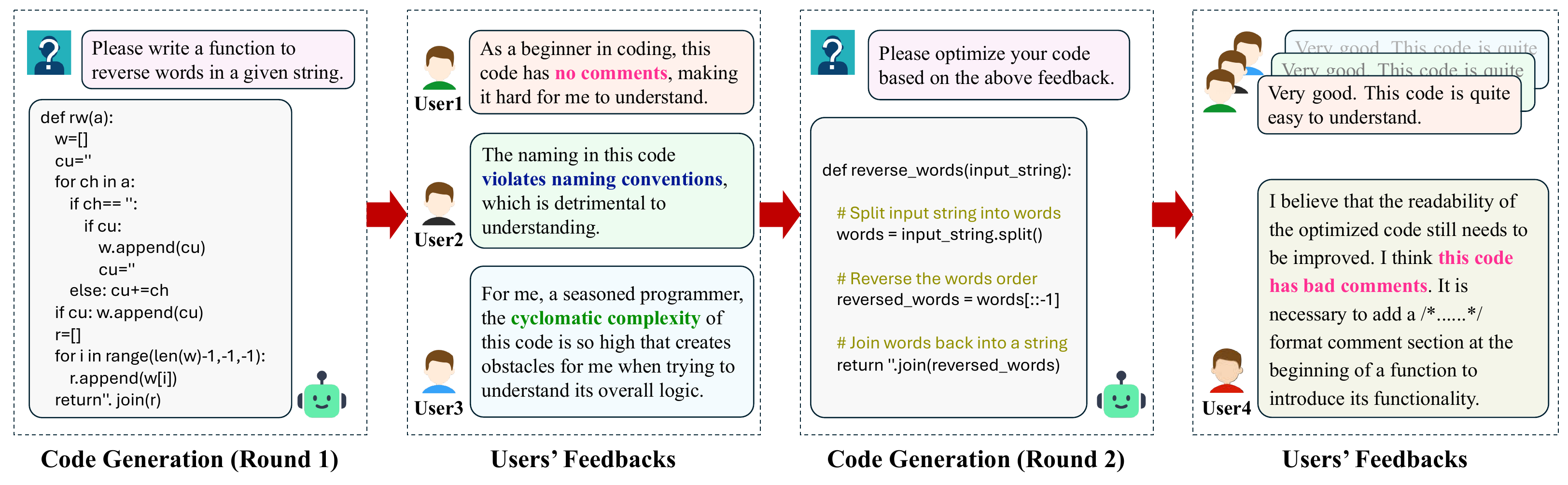}
    \caption{For different users with varying experience levels, the format of comments, naming conventions, and cyclomatic complexity can lead to differences in code readability. While User 1 (2 or 3) deems it clear and easy to understand, User 4 insists that its readability is still insufficient, advocating for the addition of a formal docstring.}
    \label{fig:Code_Readability}
\end{figure*}

As shown in~\autoref{fig:Code_Readability}, the evaluation of code readability is inherently subjective, varying across developers and experience levels. For example, for the initially generated code in the left side of~\autoref{fig:Code_Readability}, Users 1, 2, and 3 provide different critiques regarding its readability, each focusing on distinct aspects. This variation is closely related to factors such as the users' development experience and personal coding habits. Even after the code is optimized based on the requirements of the first three users, User 4 still believes that the code's readability needed further improvement. This clearly demonstrates that, given the inherently subjective nature of code readability, it is nearly impossible to evaluate it through a single, universal set of standards. Different users may expect different readability properties: novice programmers may prefer more explicit comments, whereas experienced developers may focus more on concise naming and simplified logic. This motivates a controllable method that can adapt to multiple readability preferences rather than optimizing a single static readability score.

Before the advent of large language models~(LLMs), research on code generation faced significant challenges in ensuring the code correctness~\cite{krzysztof2000generative,bengio2003neural,alon2019code2vec}. In recent years, with the advancement of LLMs, many developers have been using LLMs to generate codes on demand to support software engineering. In AI-assisted programming, LLM-generated code is often not the final artifact used in isolation. It usually needs to be reviewed, modified, debugged, and maintained by human developers. Thus, code that passes test cases may still be unsatisfactory if it lacks necessary comments, uses obscure identifiers, or contains unnecessarily complex control flow. Extensive study has been conducted on improving the LLM-generated codes' quality~\cite{wang2023review,jiang2024survey}, and have gradually recognized the importance of the code readability~\cite{takerngsaksiri2025code}. However, research aimed specifically at improving the readability of LLM-generated code remains very limited. This gap indicates that correctness and readability should be considered jointly. On the one hand, readability-oriented intervention can improve the usability and maintainability of generated code. On the other hand, such intervention should not substantially damage functional correctness, because readable but incorrect code is still unacceptable in practical software development.

Improving the readability of codes generated by LLMs can be viewed as a form of targeted control over model behavior~\cite{zhang2025style2code}. Furthermore, targeted control approaches used in real-world production environments must be capable of rapidly adapting to subjective interpretations of code readability among different users. As an approach for targeted control of model behavior, representation engineering~(RepE)~\cite{zou2023representation,wolf2025tradeoffs} achieves precise manipulation by extracting steering vectors associated with specific concepts and directly injecting them into the model's internal activation layers. Compared to other targeted control approaches~(\eg, full-parameter fine-tuning~\cite{luo2024badam,qin2024federated,lv2024full}, parameter-efficient fine-tuning~\cite{dettmers2023qlora,zhang2023adaptive,zhang2024llama}, and preference tuning~\cite{ouyang2022training,yuan2023rrhf,hong2024orpo}), RepE does not adjust model weights. Instead, it requires only a few dozen data samples targeted at a specific objective to quickly derive steering vectors that effectively control model behavior, thereby eliminating the need for large training datasets or extensive computational overhead. Given these advantages, we select RepE as the foundational approach for improving code readability in this paper. This property is suitable for readability control because different readability requirements can be represented as different steering directions, and their strengths can be adjusted during inference. As a result, RepE provides a lightweight way to transform subjective readability preferences into controllable changes in the hidden representations of LLMs, without repeatedly fine-tuning the model for each new preference.

Currently, most of the RepE work focuses on improving the model performance in a single task~(referred to as a single-task steering). However, using RepE to improve the code readability requires addressing the challenge of coordinated control across multiple metrics that constitute code readability, namely comment density, naming conventions, and cyclomatic complexity~(referred to a multitask steering). This multitask setting is more challenging than ordinary single-task steering. If steering vectors for different readability metrics are extracted independently and injected simultaneously, they may interfere with each other in the hidden representation space. For example, encouraging more comments may also affect the generated code structure, while reducing cyclomatic complexity may influence naming or decomposition choices. Therefore, improving code readability through RepE requires a mechanism that coordinates multiple steering vectors rather than simply combining them. Furthermore, when discussing how to improve the code readability, the impact on the code correctness must also be considered. Although the existing research~\cite{wolf2025tradeoffs} has explored the trade-off between the single-task steering behavior and the model capability, a unified framework for the multitask steering is still lacking~(\ie, the trade-off between the code readability and correctness in our work). In a nutshell, we aim to improve the LLM-generated code readability through RepE by addressing the following two challenges: ($a$)~How to achieve the coordinated control over the three metrics that constitute the  code readability? ($b$)~How to theoretically quantify the control impact on  the code readability and correctness, and balance the trade-off between them?

To address these challenges, we propose a multitask steering RepE framework~(MRepE) to improve the code readability, and  analyze the limit of its impact on the code correctness. Methodologically, we propose the joint principal component analysis algorithm with multidimensional orthogonal constraints~(MOC-JPCA). It introduces multidimensional orthogonal constraints and iteratively estimates the first principal direction for each dimension on the Stiefel manifold~\cite{stiefel1935richtungsfelder}, thereby explicitly suppressing mutual interference between different code readability steering vectors. This provides stable and complementary steering bases for subsequent joint injection. Compared with independently extracting steering vectors for each metric, MOC-JPCA explicitly considers the geometric relationship among different steering directions, making the extracted vectors more suitable for coordinated multitask injection. Subsequently, we analyze the functional relationship between the injection coefficients and the change in the final-layer representations, establishing two main theoretical results: a lower bound for the improvement in the code readability and an upper bound for the negative impact on the code correctness. The main contributions are:
\begin{itemize} [leftmargin=*, itemsep=0in, topsep=0in]
    \item We propose the MRepE framework to improve the code readability through multitask steering vectors.
    \item We propose the corresponding algorithm~(MOC-JPCA) to extract steering vectors in MRepE and analyze its convergence in a theoretical way.
    \item We provide both thoeretical and empirical support in the MRepE performance. Specifically, we theoretically characterize the readability-correctness trade-off and empirically validate the proposed framework on multiple code LLMs.
\end{itemize}

\section{Related Works}

Code readability has long been regarded as an important factor in software comprehension, software maintenance, and long-term code quality. Traditional studies usually attempt to relate developers' subjective perceptions of readability to measurable code-level features. For example, Buse and Weimer~\cite{buse2009learning} learned a readability metric from human annotations, while Posnett {\it et al.}~\cite{posnett2011simpler} proposed a simpler readability model based on compact textual and structural factors. Scalabrino {\it et al.}~\cite{scalabrino2017automatically} further showed that code understandability is affected by multiple source-code and documentation-related factors. These studies indicate that code readability is inherently multidimensional rather than determined by a single universal criterion. Therefore, this work focuses on three commonly recognized and practically controllable dimensions of code readability, namely comment density, naming conventions, and cyclomatic complexity.

The LLMs' application in software engineering has become increasingly widespread. Particularly, the use of LLMs for code generation has significantly enhanced productivity. However, the existing research still mainly aims to improve the correctness of LLM-generated codes~\cite{wang2023review,jiang2024survey}. For example, Bui {\it et al.}~\cite{bui2025correctness} proposed the OPENIA framework, which directly evaluates code correctness by analyzing the model's internal representations (\eg, hidden states). It outperformed traditional black-box methods on multiple benchmarks and supported more efficient quality control. This line of research is important because generated code must first satisfy functional requirements before being adopted in practical software development. Nevertheless, correctness-based evaluation mainly measures whether a generated program solves the given task, but does not fully reflect whether the program can be easily reviewed, modified, and maintained by human developers. A code snippet that passes test cases may still be difficult to understand if it lacks explanatory comments, uses obscure identifiers, or contains unnecessarily complex control flow.

This gap has motivated increasing attention to the readability of LLM-generated code. Wannita {\it et al.}~\cite{takerngsaksiri2025code} stated that, from the perspective of software engineers, the code readability remains critical in both the pre-LLMs and LLMs eras, and revealed the potential of LLMs in generating readable codes. Despite this importance, there is still no widely used quantitative method or model to assess the code readability~\cite{mannan2018towards}. In particular, the research that explicitly improves the readability of LLM-generated codes remain rather limited. Most existing studies either evaluate the readability of generated code after generation or rely on prompt engineering to encourage readable outputs. However, prompt-level control may be unstable across models and tasks, and it does not directly explain how readability-related behaviors are represented inside the model. This motivates us to actively control readability-related behaviors of LLMs through their internal representations.

Representation engineering~(RepE) has appeared as an alternative LLM tuning method. Zou {\it et al.}~\cite{zou2023representation} introduced a top-down RepE technique, such as linear artificial tomography to analyze and control high-level cognitive representations~(\eg, honesty and morality) in deep neural networks. Building on this, Wolf {\it et al.}~\cite{wolf2025tradeoffs} revealed that RepE in LLM alignment methods linearly improves attack resistance and bias control capabilities, but quadratically impairs performance on base tasks. Compared with full-parameter fine-tuning, parameter-efficient fine-tuning, and preference optimization, RepE does not update model parameters. Instead, it manipulates model behavior by extracting steering directions associated with target concepts and injecting them into hidden representations during inference. This property makes RepE suitable for code readability control, where collecting large-scale user-specific preference data for every readability style is often impractical. Moreover, the steering strength can be adjusted at inference time, which provides a flexible mechanism for balancing readability improvement and possible side effects on the original code-generation capability.

The aforementioned research on RepE primarily focuses on improving model performance on a single task. In contrast, improving code readability requires coordinated control across multiple dimensions, such as comment density, naming conventions, and cyclomatic complexity. Recent studies on activation steering also indicate that steering multiple behaviors is more challenging than steering a single behavior. For example, directly combining multiple steering vectors into one vector may lead to ineffective or unstable control, whereas injecting individual steering vectors separately can be more promising~\cite{van2024extending}. This observation is closely related to our problem setting, because each readability metric may correspond to a different representation direction in the hidden space of an LLM. If these directions are extracted independently and injected simultaneously, their geometric correlation may cause mutual interference and reduce the controllability of each metric.

To address this challenge, we propose a RepE approach for multitask steering and apply it to enhance the readability of LLM-generated codes. Specifically, we formulate the extraction of multiple readability steering vectors as a joint principal component analysis problem with multidimensional orthogonal constraints. In this way, each steering direction is encouraged to capture the dominant representation variation of its corresponding readability metric, while the orthogonality constraint explicitly suppresses interference among different metrics. Furthermore, we examine how this multitask steering influences the code correctness and present a theoretical framework that quantifies the limit of these effects. Therefore, our work differs from existing studies by jointly considering multitask readability control and the readability-correctness trade-off in LLM-generated code, rather than treating readability improvement as an isolated single-objective intervention.

\section{Improve Code Readability via MRepE}\label{sec:mrepe}
\begin{figure}[!t]
    \centering
    \includegraphics[width=1\linewidth]{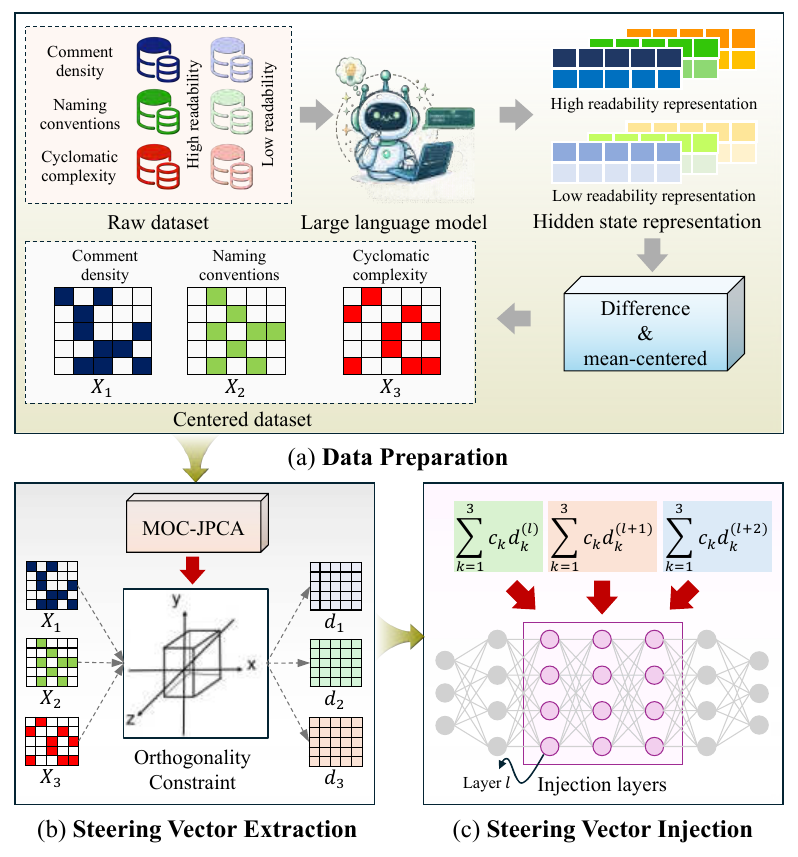}
    \caption{\yu{The MRepE framework includes three stages. ({\bf Top}) Datasets with diverse code readability metrics are provided for LLM to compute differences in hidden state representations, followed by a centering procedure. ({\bf Bottom left}) Orthogonal steering vectors are extracted using MOC-JPCA. ({\bf Bottom right}) The orthogonalized vectors are injected into selected layers of LLM to modulate and control its outputs.}}
    \label{fig:MRepE}
\end{figure}

To improve the readability of code generated by LLMs, we propose the MRepE framework as shown in Fig.~\ref{fig:MRepE}. Within this framework, we mainly focus on solutions to: ($a$)~how to use steering vectors corresponding to multiple code readability metrics to control model output; and ($b$)~how to extract these steering vectors through the MOC-JPCA algorithm. Different from weight-updating methods that require repeated optimization of the model parameters, MRepE treats readability improvement as a representation-level control problem. Specifically, it first identifies latent directions associated with different readability metrics from contrastive examples, and then injects these directions into selected hidden layers during inference. This design is particularly suitable for code readability control because different developers may emphasize different readability aspects, and such preferences should be adjusted without retraining the whole model. Therefore, the proposed framework separates the whole process into three stages: readability-oriented data preparation, orthogonal steering-vector extraction, and multitask steering-vector injection.

\subsection{Preliminary} 
We define high comment density, well-established naming conventions, and low cyclomatic complexity as the good characteristics of LLM-generated code readability. These three dimensions correspond to different levels of human code comprehension. Comment density mainly reflects whether the generated code provides sufficient natural-language explanations for important operations. Naming conventions reflect whether variables, functions, and intermediate results are expressed with meaningful and consistent identifiers. Cyclomatic complexity reflects whether the control-flow structure is simple enough for developers to trace and maintain. These dimensions represent three commonly observed sources of readability feedback: insufficient explanation, obscure naming, and unnecessarily complicated logic.

In principle, one may measure the model's relative preference by comparing the average negative log-likelihoods of code snippets with different levels of readability. However, the average negative log-likelihood of an entire code sequence is essentially a sequence-level aggregated quantity, which is jointly affected by length normalization, the proportion of attribute-irrelevant tokens, autoregressive dependence, and error accumulation. For the readability metrics considered in this work, the intervention effect may also be diluted in sequence-level averaging by a large number of tokens that are irrelevant to the target attribute. Therefore, sequence-level likelihood is more suitable as supplementary behavioral evidence, rather than as the core object of theoretical analysis in this paper. Based on this consideration, we instead use short question-answering items that can be interpreted as binary judgments~(e.g., yes/no) to test whether the model understands the indicators of highly readable code. Similarly, for code correctness, we employ multiple-choice questions to evaluate the model's ability to identify the correct code snippet.

\begin{figure}[!h]
    \centering
    \includegraphics[width=1\linewidth]{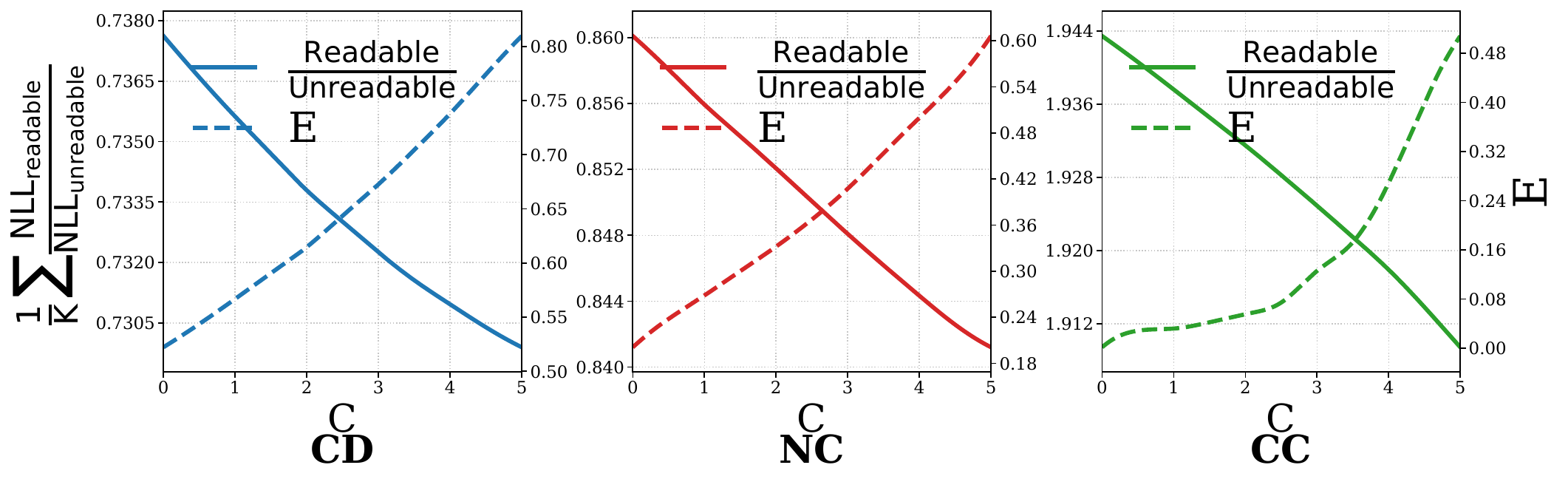}
    \caption{Injecting steering vectors for different readability metrics into Deepseek\_R1\_14b reveals a positive correlation between the model’s ability to generate readable code and its ability to understand code readability. The solid line shows the ratio of negative log‑likelihood for generating readable versus unreadable code, where a lower ratio indicates a stronger tendency to generate readable code. The dashed line represents the expected accuracy of the model’s binary judgment on code readability, where a higher value corresponds to better judgment performance.}
    \label{fig:CR_HLR_Deepseek}
\end{figure}

In addition, in our pilot experiments, we observe that after injecting steering vectors that improve the model's understanding of code readability metrics, the model's ability to generate readable code also improves, and the two are positively correlated~(as shown in Fig.~\ref{fig:CR_HLR_Deepseek}). This suggests that the judgment-based task can serve as an effective proxy for actual generation behavior. Therefore, the following analysis focuses on the probability of generating positive answers in the judgment tasks, while the sequence-level generation preference is used as behavioral evidence to verify that the learned steering directions are also reflected in actual code generation.

\subsection{Steering Vector Injection}
Given an LLM model with $\rm L$ layers for code generation, where ${\bm h}_{\theta}^{(l)}$ ($1{\le}l{\le}{\rm L}$) denotes the $l$-th layer's hidden state representation. For a given code quality-related query $q = (t_1, \cdots, t_n)$, where $t_n$ is the $n$-th token in the query, the probability distribution of the next token generated by the model with parameters $\theta$ is given by
\begin{align*}
    P_{\theta}(t_{n+1} \mid t_1, \cdots, t_n) = {\operatorname{Softmax}\left({\bm U} {\bm h}_{\theta}^{({\rm L})}(t_1, \cdots, t_n)\right)}_{t_{n+1}}
\end{align*}
where $\bm U$ is the unembedding matrix that maps the hidden state back to the space of the token vocabulary. Here, the hidden state ${\bm h}_{\theta}^{(l)}$ can be regarded as the intermediate representation through which the model encodes the semantic, syntactic, and task-related information contained in the input query. Since the final token probability is obtained by applying the unembedding matrix to the final-layer representation, a small and structured perturbation on intermediate hidden states can change the downstream answer distribution without modifying the model parameters. This is the key intuition behind steering-vector injection.

We denote queries related to comment density, naming conventions, cyclomatic complexity, and code correctness as $q_{cd}$, $q_{nc}$, $q_{cc}$, and $q_{tf}$, respectively. A positive answer $a_+$ indicates high code readability/correctness, and a negative answer $a_-$ indicates low code readability/correctness. The model's performance on the query can be quantified as the probability of generating a positive answer
\begin{align*}
    {\sum}_{a_+} P_\theta ( a_+ \mid q ) = 1 - {\sum}_{a_-} P_\theta ( a_- \mid q )
\end{align*}
For binary readability judgments, $a_+$ usually corresponds to affirmative answers such as ``yes'', while $a_-$ corresponds to negative answers such as ``no''. For multiple-choice correctness judgments, $a_+$ corresponds to the option index associated with the correct code snippet, and $a_-$ corresponds to the remaining options.

We control the output of the model defined above by injecting code readability steering vectors. The set of code readability steering vectors is defined as
\yu{
\begin{align}
    {\cal V} 
    =\left\{ {\bm v}_k^{(l)} \mid k \in \{ 1, 2, \cdots, {\rm K} \} , l \in \{ 1, 2, \cdots, {\rm  L} \} \right\}
    \label{eq:steering_vector}
\end{align}
In~\autoref{eq:steering_vector}, ${\bm v}_k^{(l)}{=}c_k{\cdot}{\bm d}_k^{(l)}$, where $c_k$ is the coefficient specifying steering strength, and ${\bm d}_k^{(l)}$ denotes steering direction.} $\rm K$ is the number of code readability steering vectors~(${\rm K}=3$ in this paper). \yu{For layers that are not injected with code readability steering vectors, we set $\|{\bm d}_k^{(l)}\| = 0$}. The coefficient $c_k$ determines how strongly the model is encouraged to follow the $k$-th readability direction. In this paper, the three directions correspond to comment density, naming conventions, and cyclomatic complexity. A larger coefficient imposes a stronger representation-level bias toward the corresponding readability metric, while a smaller coefficient leads to a more conservative intervention. Therefore, the coefficient vector ${\bm c}=(c_1,\cdots,c_{\rm K})$ provides a direct interface for balancing the improvement of different readability dimensions.

The model injected with the readability steering vectors is denoted as $P_{\theta , {\cal V}}$. The steering vectors are applied by modifying the hidden state representation at each layer as
\begin{align}
    {\bm h}_{\theta{},{\bm v}^{(l)}}^{(l)} \leftarrow {\bm h}_\theta{}^{(l)} + {\bm v}_1^{(l)} + \cdots + {\bm v}_{\rm  K}^{(l)}
    \label{eq:vector_injection}
\end{align}
The additive form in~\autoref{eq:vector_injection} is simple but important. It implies that different readability preferences are jointly expressed as a superposition of several steering directions in the hidden space. If these directions are highly correlated, the injected perturbations may reinforce or cancel each other in an unpredictable way. Therefore, simply extracting each steering direction independently is insufficient for stable multitask control, which motivates the orthogonal extraction algorithm introduced later. The probability that the model $P_{\theta , {\cal V}}$ generates a positive answer is
\begin{align*}
    {\sum}_{a_+} P_{\theta, { \cal V }} ( a_+ \mid q ) = 1 - {\sum}_{a_-} P_{\theta, { \cal V }} ( a_- \mid q )
\end{align*}

By comparing ${\sum}_{a_+} P_{\theta, { \cal V }} ( a_+ \mid q )$ and ${\sum}_{a_+} P_\theta ( a_+ \mid q )$, we can analyze the impact of the code readability steering vectors on both code readability and correctness. Specifically, an increase in ${\sum}_{a_+} P_{\theta, { \cal V }} ( a_+ \mid q_{cd})$, ${\sum}_{a_+} P_{\theta, { \cal V }} ( a_+ \mid q_{nc})$, or ${\sum}_{a_+} P_{\theta, { \cal V }} ( a_+ \mid q_{cc})$ indicates that the model becomes more sensitive to the corresponding readability metric. In contrast, a decrease in ${\sum}_{a_+} P_{\theta, { \cal V }} ( a_+ \mid q_{tf})$ indicates a possible degradation in the model's correctness-related judgment ability. Hence, the same injection mechanism provides a unified way to quantify both the desired readability improvement and the possible side effect on correctness.

\subsection{Data Preparation for Readability Representations}

Before extracting steering vectors, MRepE constructs contrastive representation datasets for different readability metrics. For each metric $k$, we build a set of paired prompts
\begin{align}
    \{(x_{k,i}^{+}, x_{k,i}^{-})\}_{i=1}^{{\rm M}_k}
\end{align}
where $x_{k,i}^{+}$ denotes a code-related input that better satisfies the $k$-th readability metric, and $x_{k,i}^{-}$ denotes its low-readability counterpart. For example, for the comment-density metric, $x_{k,i}^{+}$ contains sufficient explanatory comments, whereas $x_{k,i}^{-}$ contains insufficient or missing comments. For the naming-convention metric, the positive example uses meaningful and conventional identifiers, whereas the negative example uses obscure or inconsistent identifiers. For the cyclomatic-complexity metric, the positive example presents a simpler control-flow structure, whereas the negative example contains unnecessarily complex branches or loops.

For each pair $(x_{k,i}^{+},x_{k,i}^{-})$, we feed the two inputs into the LLM and obtain the hidden representations at a selected layer $l$. The representation difference is computed as
\begin{align}
    \Delta {\bm h}_{k,i}^{(l)}
    =
    {\bm h}_{\theta}^{(l)}(x_{k,i}^{+})
    -
    {\bm h}_{\theta}^{(l)}(x_{k,i}^{-})
    \label{eq:representation_difference}
\end{align}
The difference vector in~\autoref{eq:representation_difference} captures the local representation shift from a low-readability example to a high-readability example under the same metric. We then mean-center these difference vectors to remove the common offset shared by all pairs
\begin{align}
    {\bm x}_{k,i}^{(l)}
    =
    \Delta {\bm h}_{k,i}^{(l)}
    -
    \frac{1}{{\rm M}_k}\sum_{j=1}^{{\rm M}_k}\Delta {\bm h}_{k,j}^{(l)}
    \label{eq:mean_centering}
\end{align}
The centered vectors are stacked to form the dataset $X_k^{(l)}=[{\bm x}_{k,1}^{(l)},\cdots,{\bm x}_{k,{\rm M}_k}^{(l)}]^{\top}$. In the following optimization problem, we omit the superscript $l$ for notational simplicity, since the same extraction procedure can be independently applied to each selected injection layer.

\subsection{Steering Vector Extraction}

\zeng{To improve the readability of code generated by LLMs, we need to extract steering vectors associated with code readability metrics, including comment density, naming conventions, and cyclomatic complexity. As shown in Eq.~\ref{eq:vector_injection}, these steering vectors are injected into the model’s internal activation layers, enabling targeted control over the code readability metrics by modulating the hidden state representations of the model.} A natural baseline is to extract the steering vectors independently for each code readability metric through PCA, and inject them into the model simultaneously. However, the resulting vectors exhibit substantial and irregular mutual interference~\cite{van2024extending}. The reason is that independent PCA only maximizes the projected variance within each individual dataset and does not consider the geometric relationship among different readability directions. Consequently, two steering vectors extracted from different readability metrics may still point to highly overlapping subspaces. When such correlated vectors are injected together, the actual representation perturbation is no longer metric-specific: one vector may unintentionally affect another readability metric or even amplify the disturbance to code correctness. This phenomenon is especially problematic in multitask readability control, where the purpose is not merely to improve one metric, but to obtain controllable and complementary improvements across several metrics. To address this challenge, we need to solve an optimization problem that jointly estimates the first principal component for each dataset under the constraint that all principal components are pairwise orthogonal. The formal definition of this optimization problem is as follows.

\yu{Given $\rm K$ groups of centered datasets $\{X_1, X_ 2\cdots , X_{\rm K}\}$, where each dataset $X_k \in \mathbb{R}^{{\rm M}_k{\times}{\rm N}}$ corresponds to a code readability metric, ${\rm M}_k$ is the number of samples in the $k$-th dataset, and $\rm N$ is the feature dimension of each sample}, the objective is to find $\rm K$ unit vectors $\{{\bm d}_1 , {\bm d}_2 , \cdots , {\bm d}_{\rm K}\}$ that maximize the weighted sum of the projected variance
\begin{equation}
\label{op:moc-jpca}
\begin{aligned}
\max_{{\bm d}_1 , {\bm d}_2 , \cdots , {\bm d}_{\rm K}} & \sum_{k=1}^{\rm K} {\bm d}_{k}^{\top} {\bm S}_k {\bm d}_k \\
{\rm s.t.} & \ {\bm D}^{\top} {\bm D} = {\bm I}
\end{aligned}
\end{equation}
where $\boldsymbol{S}_{k} = \frac{1}{{\rm M}_{k} - 1} X_{k}^{\top} X_{k}$ is the covariance matrix of \yu{the $k$-th dataset}, ${\bm D} = \left[ {\bm d}_1 , {\bm d}_2 , \cdots , {\bm d}_{\rm K} \right]$ is the matrix of unit vectors, and $\bm I$ is the identity matrix, enforcing orthogonality among the ${\bm d}_k$ vectors. ${\cal M} = \left\{ {\bm D} \in \mathbb{R}^{{\rm N} \times {\rm K}} \mid {\bm D}^{\top} {\bm D} = {\bm I} \right\}$ is a Stiefel manifold. The constraint ${\bm D}^{\top}{\bm D}={\bm I}$ has two effects. First, it normalizes each steering direction, which makes the coefficient $c_k$ in~\autoref{eq:steering_vector} comparable across different readability metrics. Second, it enforces ${\bm d}_i^{\top}{\bm d}_j=0$ for $i\neq j$, which explicitly reduces the overlap among different readability directions.

From the perspective of multitask control, the objective term ${\bm d}_{k}^{\top}{\bm S}_k{\bm d}_k$ encourages ${\bm d}_{k}$ to preserve the most informative representation variation associated with the $k$-th readability metric. The orthogonality constraint, on the other hand, prevents the solution from assigning similar semantic effects to multiple steering directions. Therefore, MOC-JPCA can be viewed as a compromise between metric-specific expressiveness and cross-metric disentanglement. This is different from applying Gram-Schmidt orthogonalization after independent PCA, because post-processing may rotate a direction away from the high-variance subspace of its own dataset, whereas MOC-JPCA optimizes the variance objective and the orthogonality constraint simultaneously.

To solve the optimization problem formulated in~(\ref{op:moc-jpca}) above, we develop the joint principal component analysis algorithm with multidimensional orthogonal constraints~(MOC-JPCA) in Algorithm~\ref{alg:moc_jpca}. Specifically, given ${\rm K}$ groups of centered datasets, we first calculate the covariance matrix ${\bm S}_k$ for each dataset~(lines 1--3). Then, we initialize the unit vector matrix ${\bm D}$ in an orthogonal space~(line 4). After that, we iteratively update ${\bm D}$ on the Stiefel manifold. In each iteration, we first compute the covariance product matrix ${\bm B}$~(line 6), and then obtain the Riemannian gradient ${\bm G}$ by projecting the Euclidean gradient onto the tangent space of the Stiefel manifold~(line 7). To ensure a sufficient increase of the objective function, the update step size $\eta$ is determined by the Armijo backtracking line search~(lines 8--11). Finally, the updated matrix is retracted back onto the Stiefel manifold via the QR retraction~(line 12). The loop is executed for a maximum of $\rm T$ iterations. More concretely, the matrix ${\bm B}=[{\bm S}_1{\bm d}_1,\cdots,{\bm S}_{\rm K}{\bm d}_{\rm K}]$ collects the Euclidean ascent directions for all readability metrics. Since a direct Euclidean update may violate the orthogonality constraint, the algorithm projects this direction onto the tangent space of the Stiefel manifold. The projection term ${\bm D}\operatorname{sym}({\bm D}^{\top}{\bm B})$ removes the component that would move ${\bm D}$ outside the feasible manifold. After taking a step along the projected direction, the QR-based mapping $\operatorname{qf}(\cdot)$ retracts the updated matrix back to a feasible orthogonal matrix. In this way, every iterate satisfies ${\bm D}^{\top}{\bm D}={\bm I}$, and the extracted directions can be directly used as steering directions. The following theorem establishes the convergence of the MOC-JPCA algorithm, and the proof details are provided in Appendix C-A.

\begin{theorem}\label{theorem3}
\begingroup\itshape
Let ${\bm D}^{(0)}$ be randomly generated from an orthogonal space, and let $\{{\bm D}^{(t)}\}$ be the iterative sequence produced by the MOC-JPCA algorithm, where ${\bm D}^{(t)} \in \mathbb{R}^{N \times K}$ denotes the solution at the $t$-th iteration. Then, the sequence $\{{\bm D}^{(t)}\}$ converges to a first-order stationary point of Problem~(3).
\endgroup
\end{theorem}

Theorem~\ref{theorem3} provides the theoretical basis for using MOC-JPCA as the extraction module of MRepE. Since the objective in~\eqref{op:moc-jpca} is generally non-convex on the Stiefel manifold, the theorem does not claim global optimality. Instead, it ensures that the iterative process converges to a point at which no first-order feasible ascent direction exists. This guarantee is sufficient for the present framework because the extracted steering vectors are used as representation-level control directions rather than as exact global solutions of a convex problem.

\begin{algorithm}[!htbp]
    \caption{The joint principal component analysis with multidimensional orthogonal constraints~(MOC-JPCA)}
    \renewcommand{\algorithmicrequire}{\textbf{Input:}}
    \renewcommand{\algorithmicensure}{\textbf{Output:}}
    \begin{algorithmic}[1]
    \REQUIRE The centered datasets $\{X_1, X_2, \cdots, X_{\rm K}\}$; parameters $\bar{\eta}, \beta, c$
    \ENSURE The set of unit vectors $\{{\bm d}_1, {\bm d}_2, \cdots, {\bm d}_{\rm K}\}$
        \FOR{$k = 1, 2, \cdots, {\rm K}$}
            \STATE{$\boldsymbol{S}_{k} \leftarrow \frac{1}{{\rm M}_{k} - 1} X_{k}^{\top} X_{k}$;}
        \ENDFOR
        \STATE{Initialize $\bm D = \left[ {\bm d}_1 , {\bm d}_2 , \cdots , {\bm d}_{\rm K} \right]$ such that $\bm D^\top \bm D = \bm I$;}
        \FOR{$t = 1, 2, \cdots, {\rm T}$}
            \STATE{$\bm B \leftarrow \left[ {\bm S}_{1}{\bm d}_{1}, \cdots, {\bm S}_{\rm K}{\bm d}_{\rm K} \right]$;}
            \STATE{$\bm G \leftarrow 2\left(\bm B - \bm D \operatorname{sym}(\bm D^\top \bm B)\right)$;}
            \STATE{$\eta \leftarrow \bar{\eta}$;}
            \WHILE{$f(\operatorname{qf}(\bm D + \eta \bm G)) < f(\bm D) + c \eta \|\bm G\|_F^2$}
                \STATE{$\eta \leftarrow \beta \eta$;}
            \ENDWHILE
            \STATE{$\bm D \leftarrow \operatorname{qf}(\bm D + \eta \bm G)$;}
        \ENDFOR
    \end{algorithmic}
    \label{alg:moc_jpca}
\end{algorithm}

After Algorithm~\ref{alg:moc_jpca} returns the orthogonal direction matrix ${\bm D}$, each column ${\bm d}_{k}$ is assigned to the corresponding readability metric and scaled by its coefficient $c_k$ during inference. In the full MRepE pipeline, the extraction step and the injection step are decoupled: the directions are computed once from the contrastive readability datasets, whereas the coefficients can be adjusted at inference time according to the desired trade-off between readability and correctness. This decoupling makes MRepE flexible in practical scenarios. For example, when the user requires more explanatory comments, the coefficient of the comment-density direction can be increased; when correctness is more important, all coefficients can be kept small to reduce the intervention strength.

\section{Theoretical Results}\label{sec:theoretical_analysis}
 As shown in empirical study, the improvement of code readability might come at the cost of reduced code correctness. By controlling the steering strength of the vectors across different metrics, the model can achieve balanced performance in the code readability and correctness. Specifically, we conduct this analysis in a theoretical way. Theorem~\ref{theorem1} shows, if a query corresponds to any metric included among the injected code readability metrics, the model's capability on that metric is guaranteed with a lower bound. Theorem~\ref{theorem2} establishes that if the query corresponds to the code correctness, the model's capability regarding the code correctness is constrained by an upper bound. In other words, the negative impact on the code correctness due to the improved readability is to be bounded. {The detailed proof is presented in Appendixes C-B and C-C.}

\begin{theorem}\label{theorem1}
\begingroup\itshape
Let $P_{\theta, {\cal V}}({\cdot}{\mid}q)$ denote a model \yu{that is injected with a set of code readability steering vectors $\{{\cal V}_1, \cdots, {\cal V}_{\rm K}\}$}, where the query $q$ is uniquely associated with one of the steering vectors. The model's performance on the code readability metric corresponding to query $q$ is evaluated by the probability of generating a positive answer $\sum_{a_+} P_{\theta,{ \cal V}} (a_+{\mid}q)$. Let $\delta_{\cal V}(q) = A\left(1-e^{-\sqrt{\sum_{k} \lambda_{k}c^2_k}}\right)$ denote the change in the final hidden layer representation induced by the injected steering vectors. Then, the model’s performance on the code readability metric corresponding to query $q$ is lower-bounded by
\endgroup
\begin{align*}
\sum{}_{a_{+}} P_{\theta, { \cal V }}\left(a_{+} \mid q\right) \ge \sigma\left(\sigma^{-1}\left(P_{0}\right) + \Delta_1 \cdot \delta{}_{ \cal V } ( q ) + \Delta_2 \right)
\end{align*}
\end{theorem}

Here, \zeng{$\sigma(x)=\frac{1}{1+e^{-x}}$ is the logistic function and } $P_0$ is the expected probability of the model generating a positive answer to query $q$ without any steering vector intervention. $A$ and $\lambda_k$ are model-dependent and characterize the relationship between $c_k$ and the norm of the corresponding final hidden layer representation. $\Delta_1 > 0$ indicates the performance improvement \yu{through} the code readability steering vector, and $\Delta_2$ indicates the potential negative impact caused by steering vectors irrelevant to the query $q$. Since this work does not consider impairments to code readability, all steering strength coefficients are constrained to satisfy $c_k > 0 (1 \le k \le {\rm K})$.

As evident from the mathematical expression, this lower bound is influenced by all steering vector coefficients $c_k (1 \le k \le {\rm K})$. At $\sqrt{ \sum_{1 \leq k \leq {\rm K}} \lambda{}_{k} c^2_k } \rightarrow 0$, the lower bound reduces to $P_0 + \sigma( \Delta_2 )$, approximating the expected probability of the model generating a positive answer without any steering vector injection. At $\sqrt{ \sum_{1 \leq k \leq {\rm K}} \lambda{}_{k} c^2_k } \rightarrow \infty$, the lower bound approaches $\sigma ( \sigma^{-1}\left(P_{0}\right)+\Delta_1 \cdot A +\Delta_2 )$.

\begin{theorem}\label{theorem2}
\begingroup\itshape
Let $P_{\theta , { \cal V } } ( \cdot \mid q )$ denote a model injected with code readability steering vectors ${ \cal V }_1 , ... , { \cal V }_{\rm K}$, where the query $q$ is associated with code correctness. Assume that, for the representations of positive and negative answers constituting a probability mass of $1 - \epsilon$, the change in the final hidden layer representation induced by the injected steering vectors follows a random distribution with variance ${\sigma}^2 > 0$. Then, with probability at least $1 - \frac{2}{\rm T}$, the probability of the model generating a positive answer to query $q$ is upper-bounded by
\endgroup
\begin{align*}
P_{\theta, { \cal V } }\left(a_{+} \mid q\right) \leq \frac{ P_0 }{ P_0 + ( 1 - P_0 ) \alpha ( 1 - \epsilon ) ( 1 + \frac{1}{2} {\beta}^2 {\sigma}^2 { \delta{}_{ \cal V } ( q ) }^2 )}
\end{align*}
\end{theorem}

Here, $\rm T$ denotes the number of tokens constituting the probability mass $1 - \epsilon$, and $\alpha$ and $\beta$ are parameters dependent on query $q$.

This theorem characterizes the effect of injecting the code readability steering vectors on tasks of code correctness. 
The parameter $\alpha \in \left[ 0 , 1 \right]$ quantifies the looseness of the bound caused by asymmetry in the probability distribution. Specifically, at $\sqrt{ \sum_{1 \leq k \leq {\rm K}} \lambda{}_{k} c^2_k } \rightarrow 0$, the true probability of the model generating a positive answer to a code correctness-related query is $P_0$, while the upper bound given by the theorem becomes $\frac{ P_0 }{ P_0 + ( 1 - P_0 ) \alpha ( 1 - \epsilon ) }$. Consequently, at $\alpha = 1$ and $\epsilon = 0$, the upper bound at $\sqrt{ \sum_{1 \leq k \leq {\rm K}} \lambda{}_{k} c^2_k } = 0$ exactly coincides with the true probability. A small value of $\alpha$ indicates that the bound overestimates the true probability.

The coefficients $\lambda_{k}$ capture how the code readability steering vectors influence the code correctness. The product $\beta \sigma$ characterizes the probability decay rate, where $\sigma$ is the standard deviation of random noises imposed on the correctness due to the injection of the code readability steering vectors, and $\beta$ is the minimum of the two weighted sums of positive and negative random variables parameterized with $\sigma^\prime = 1$.

\section{Experimental Results}\label{sec:experiments}
We evaluate the code readability and correctness by examining how they vary with increasing coefficients of the code readability steering vectors. \yu{Our aim is to} verify the aforementioned theoretical bounds, specifically to \yu{($a$) demonstrate the performance improvement of the code readability via the injected steering vectors; and ($b$) discuss the impact of the injected steering vectors on code correctness}.

\subsection{Experimental Setup}\label{subsec:experiments_Setup}
\yu{For ease of notation, we let} comment density, naming conventions, cyclomatic complexity, and code correctness be $\rm CD$, $\rm NC$, $\rm CC$, and $\rm TF$, respectively. The steering strength associated with a metric is denoted as ${\rm c}(\cdot)$, and the average probability of generating positive answers related to a metric is denoted as ${\rm E}(\cdot)$.

\paragraph{Datasets} \yu{We use the mostly basic Python problems (MBPP)~\cite{austin2021program} as a benchmark dataset to evaluate the quality of LLM-generated codes. The benchmark consists of around $1,000$ crowd-sourced Python programming problems.} We extend $300$ samples from the MBPP dataset into two new datasets designed to evaluate code readability and code correctness of code generated by LLMs. The extension format for assessing metrics such as comment density, naming conventions, and cyclomatic complexity is illustrated in the following example~(referred to as MBPP-CR dataset, Tbl.~\ref{tab:sample_mbpp_cr}).
\begin{table}[ht]
    \centering
    \caption{An example from the MBPP-CR dataset}
    \resizebox{\linewidth}{!}{
    \begin{tabular}{|c|c|c|}
    \hline
    Question & CD\_LOW & CD\_HIGH \\
    \hline
    \makecell[l]{Write a function\\ to reverse words\\ in a given string.} &  \makecell[l]{def reverse\_words(s):\\ \quad{}\quad{}return ' '.join(reversed(s.split()))} & \makecell[l]{def reverse\_words(s):\\ \quad{}\quad{}\#Split string into words,\\ \quad{}\quad{}reverse their order,\\ \quad{}\quad{}and join back with spaces\\ \quad{}\quad{}return ' '.join(reversed(s.split()))} \\
    \hline
    \multicolumn{1}{c}{} & \multicolumn{1}{c}{} & \multicolumn{1}{c}{} \\
    \hline
    Question & NC\_LOW & NC\_HIGH \\
    \hline
    \makecell[l]{Write a function\\ to reverse words\\ in a given string.} &  \makecell[l]{def r\_w(s):\\ \quad{}\quad{}return ' '.join(reversed(s.split()))} & \makecell[l]{def reverse\_words(s):\\ \quad{}\quad{}return ' '.join(reversed(s.split()))} \\
    \hline
    \multicolumn{1}{c}{} & \multicolumn{1}{c}{} & \multicolumn{1}{c}{} \\
    \hline
    Question & CC\_LOW & CC\_HIGH \\
    \hline
    \makecell[l]{Write a function\\ to reverse words\\ in a given string.} &  \makecell[l]{def reverse\_words(s):\\ \quad{}\quad{}return ' '.join(reversed(s.split()))} & \makecell[l]{def reverse\_words(s):\\ \quad{}\quad{}words = []\\ \quad{}\quad{}for word in s.split():\\ \quad{}\quad{}\quad{}\quad{}words.insert(0, word)\\ \quad{}\quad{}return ' '.join(words)} \\
    \hline
    \end{tabular}
    }
    \label{tab:sample_mbpp_cr}
\end{table}
The extension format for evaluating correctness is as follows~(referred to as MBPP-CC dataset, Tbl.~\ref{tab:sample_mbpp_cc}).
\begin{table}[ht]
    \centering
    \caption{An example from the MBPP-CC dataset}
    \resizebox{\linewidth}{!}{
    \begin{tabular}{|c|c|c|c|c|c|}
    \hline
    Question & A & B & C & D & Answer\\
    \hline
    \makecell[l]{Write a function\\ to reverse words\\ in a given string.} &  \makecell[l]{def reverse\_words(s: str) $\rightarrow$ str:\\ \quad{}\quad{}return s[::-1]} & \makecell[l]{def reverse\_words(s: str) $\rightarrow$ str:\\ \quad{}\quad{}words = s.split()\\ \quad{}\quad{}reversed\_word\_iterator = reversed(words)\\ \quad{}\quad{}return ' '.join(reversed\_word\_iterator)} & ...... & ...... & B\\
    \hline
    \end{tabular}
    }
    \label{tab:sample_mbpp_cc}
\end{table}

\paragraph{Query Templates} For the evaluation queries, we construct four types of query templates corresponding to comment density, naming conventions, cyclomatic complexity, and code correctness, denoted as $q_{cd}$, $q_{nc}$, $q_{cc}$, and $q_{tf}$, respectively. For $q_{cd}$, $q_{nc}$, and $q_{cc}$, the positive labels correspond to high comment density, good naming conventions, and low cyclomatic complexity, respectively. For $q_{tf}$, the positive label corresponds to the correct code option. The query templates are as follows
\begin{itemize} [leftmargin=*, itemsep=0in, topsep=0in]
    \item Template of $q_{cd}$: \\
    ``Is the comment density of the following Python code high/low? \\
    \{code\}''
    
    \item Template of $q_{nc}$: \\
    ``Do you agree that the following code has good/bad naming conventions? \\
    \{code\}''
    
    \item Template of $q_{cc}$: \\
    ``Is the cyclomatic complexity of the following Python code low/high? \\
    \{code\}''
    
    \item Template of $q_{tf}$: \\
    ``\{question\} \\
    For the programming problem above, which one of the following four Python code options (A, B, C, D) is correct? \\
    A) \{codeA\} \\
    B) \{codeB\} \\
    C) \{codeC\} \\
    D) \{codeD\}''
\end{itemize}

\paragraph{Foundation Models} We use three foundation language models: ($a$) \textbf{Deepseek\_R1\_14b} \cite{guo2025deepseek} is an efficient reasoning model obtained through knowledge distillation from DeepSeek-R1. ($b$) \textbf{Qwen2.5coder\_14b\_Instruct} \cite{hui2024qwen2} is a 14.7-billion-parameter causal language model instruction-aligned for programming tasks by Tongyi Qianwen. ($c$) \textbf{Codellama\_13b\_Instruct} \cite{roziere2023code} is a 13-billion-parameter code generation model built upon the Llama2 architecture.

\paragraph{Experimental Equipments} All the experiments are conducted on two computing servers \yu{({\bf S1} and {\bf S2}). S1 runs on Ubuntu 22.04.3 LTS and is equipped with a dual Intel Xeon Gold 6530 CPU, 512GB RAM, and 4 NVIDIA GeForce RTX 4090 GPUs. S2 operates on Ubuntu 22.04.4 LTS and featured with a dual Intel(R) Xeon(R) Gold 6248R CPU, 1.0TB RAM, and 4 NVIDIA Tesla V100 SXM3 GPUs.}

\subsection{Steering Vector Extraction}\label{subsec:steering_vector_extraction}
\begin{figure}[!htbp]
    \centering
    % 加载 floatrow 后，可以使用 \sidesubfloat 将标签置于左侧
    \subfloat[]{
        \includegraphics[width=0.9\linewidth]{figs/Deepseek_14b_Instruct_CD_NC_CC_high_low_ratio_with_readable_prob.pdf}
        \label{fig:cr_jpca_m1}
    }
    \\
    \subfloat[]{
        \includegraphics[width=0.9\linewidth]{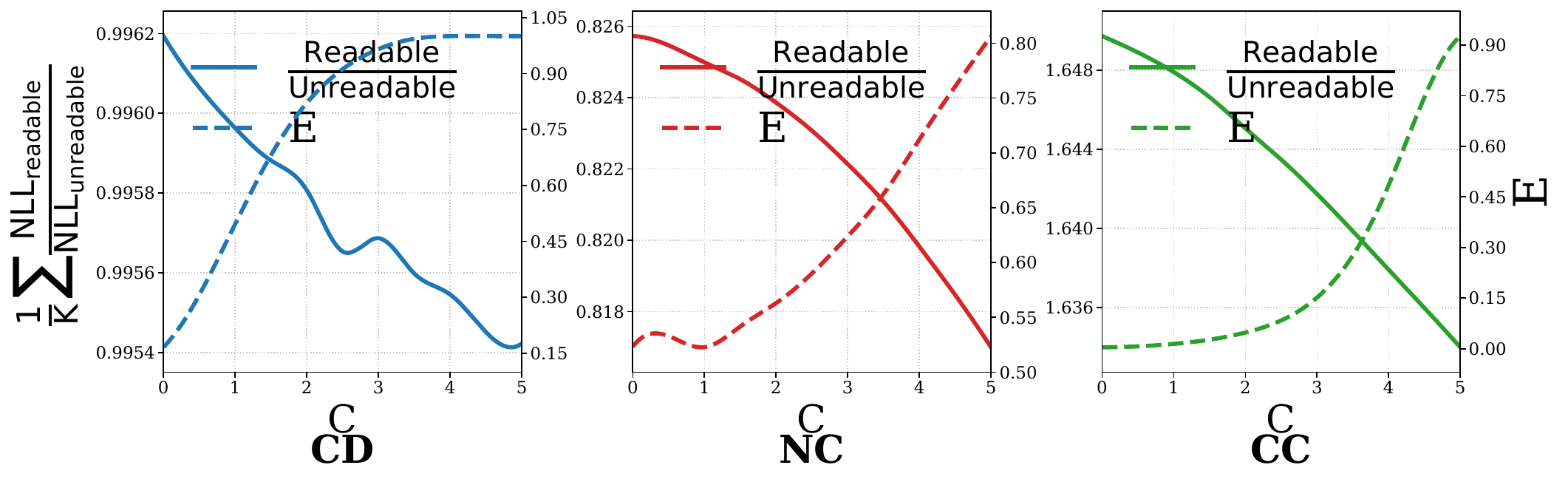}
        \label{fig:cr_jpca_m3}
    }
    \\
    \subfloat[]{
        \includegraphics[width=0.9\linewidth]{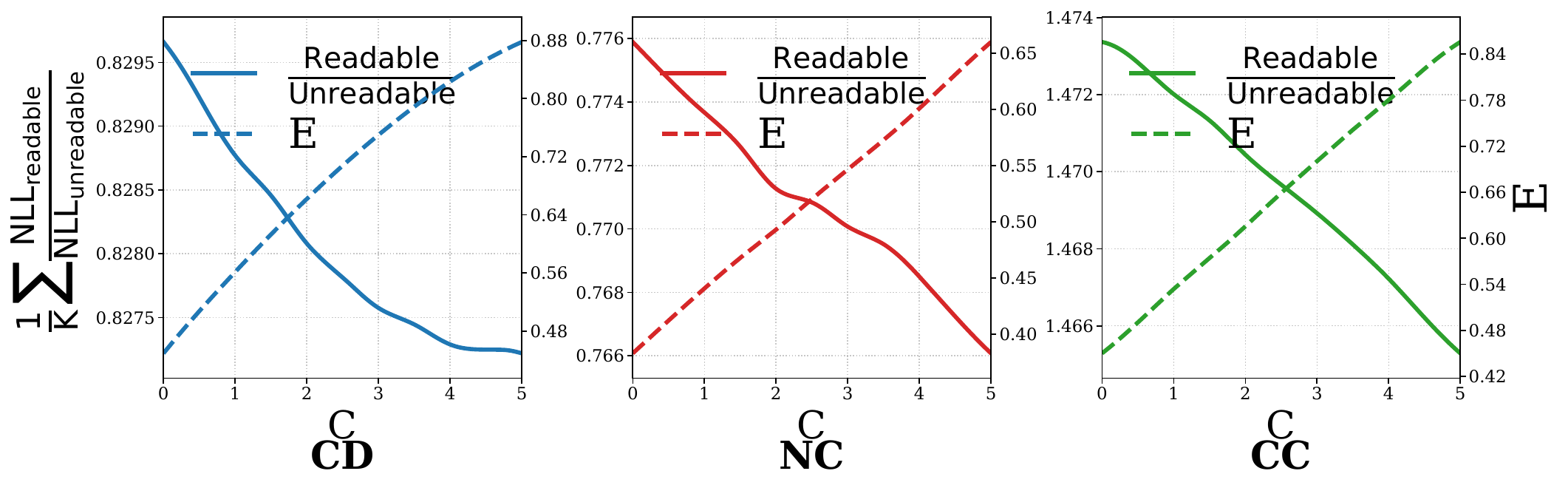}
        \label{fig:cr_jpca_m4}
    }
    \caption{The readability performance for the models with the single-metric varying with steering strength.}
    \label{fig:cr_jpca}
\end{figure}

\yu{We first evaluate the effectiveness of the steering vector extraction based on MOC-JPCA on the MBPP-CR dataset. \autoref{fig:cr_jpca} shows the probability that}, \zeng{after injecting the corresponding steering vector at varying strengths~(${\rm c}(\cdot)$) for a given code readability metric~(either CD, NC or CC), we report the ratio of the average negative log-likelihood for generating readable versus unreadable code~($\frac{1}{\rm K} \sum \frac{NLL_{readable}}{NLL_{unreadable}}$), and the probability of the model providing a positive answer~(${\rm E}$) to queries related to that metric.} {Details on the specific layers into which each steering vector is injected and the significance analysis of the average negative log-likelihood ratio are provided in Appendix D-A.} {The model performance improves as the steering strength increases, which validates the effectiveness of the steering vectors extracted by MOC-JPCA .} 

\subsection{Code Readability Measurement}\label{subsec:code_readability_measurement_2sv}
\begin{figure}[!t]
    \centering
    % 第一行：Deepseek_R1_14b模型
    \subfloat[Results of Deepseek\_R1\_14b model]{
    \begin{minipage}[t]{0.99\linewidth}
    \centering
    \includegraphics[width=.32\linewidth]{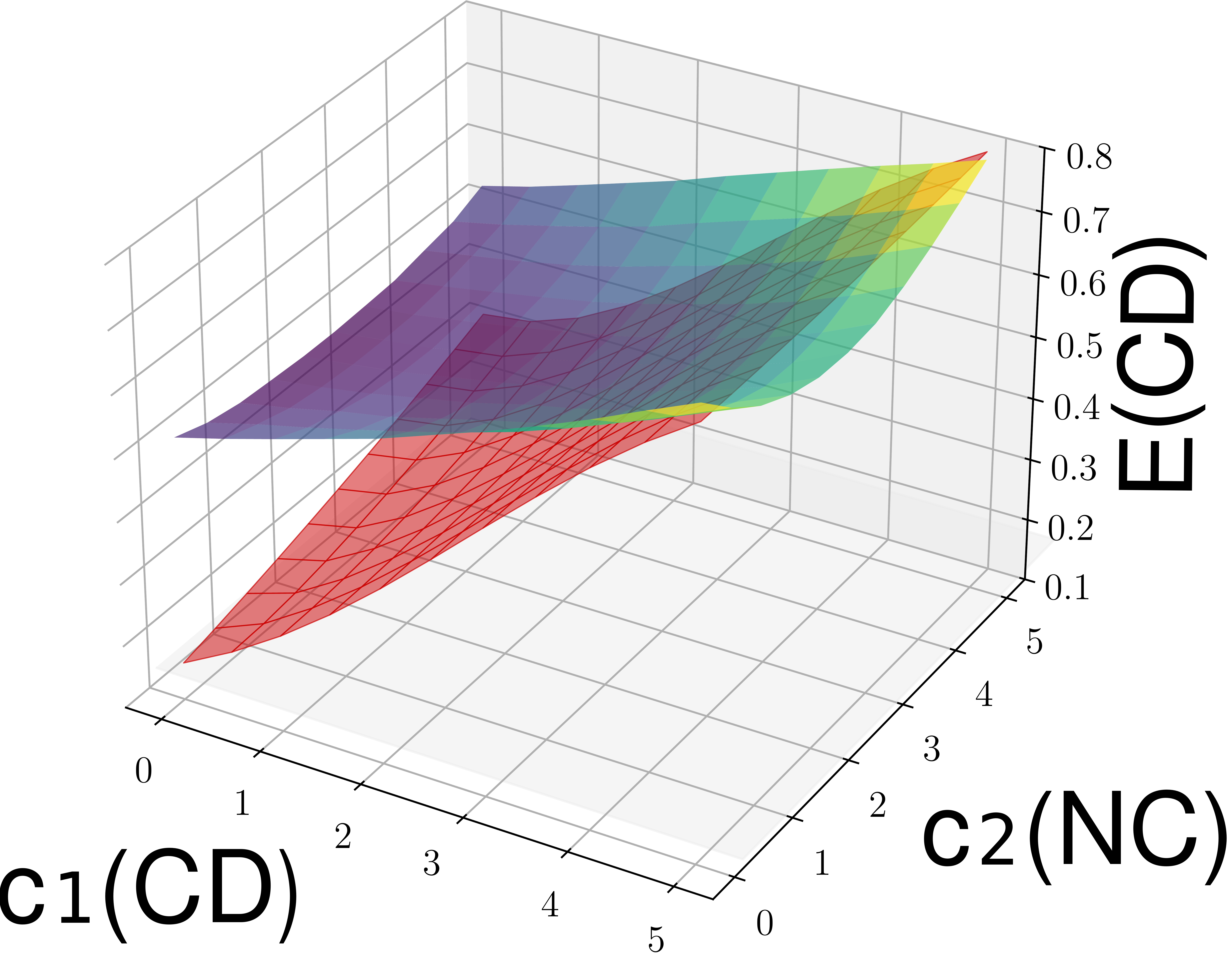}
    \includegraphics[width=.32\linewidth]{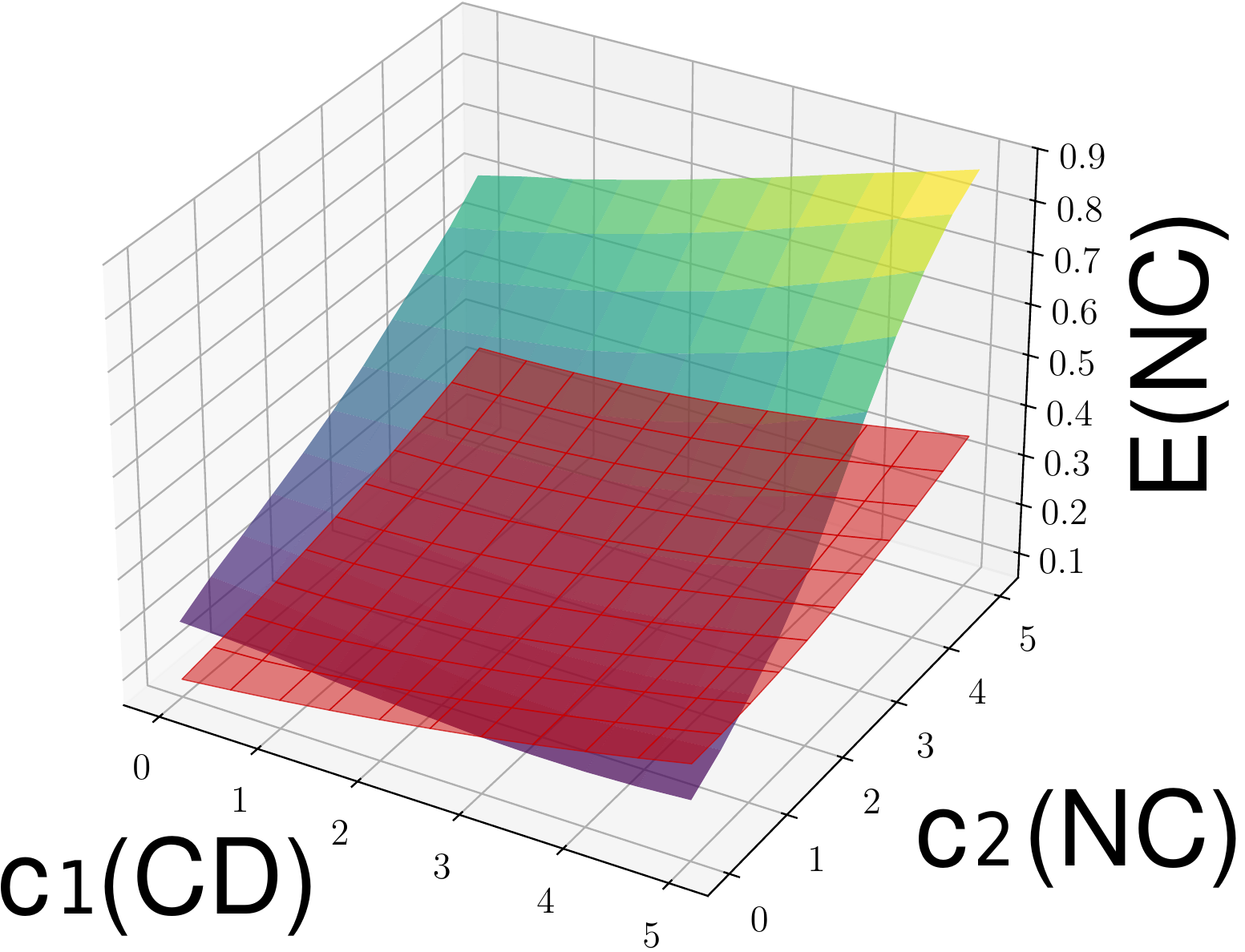}
    \includegraphics[width=.32\linewidth]{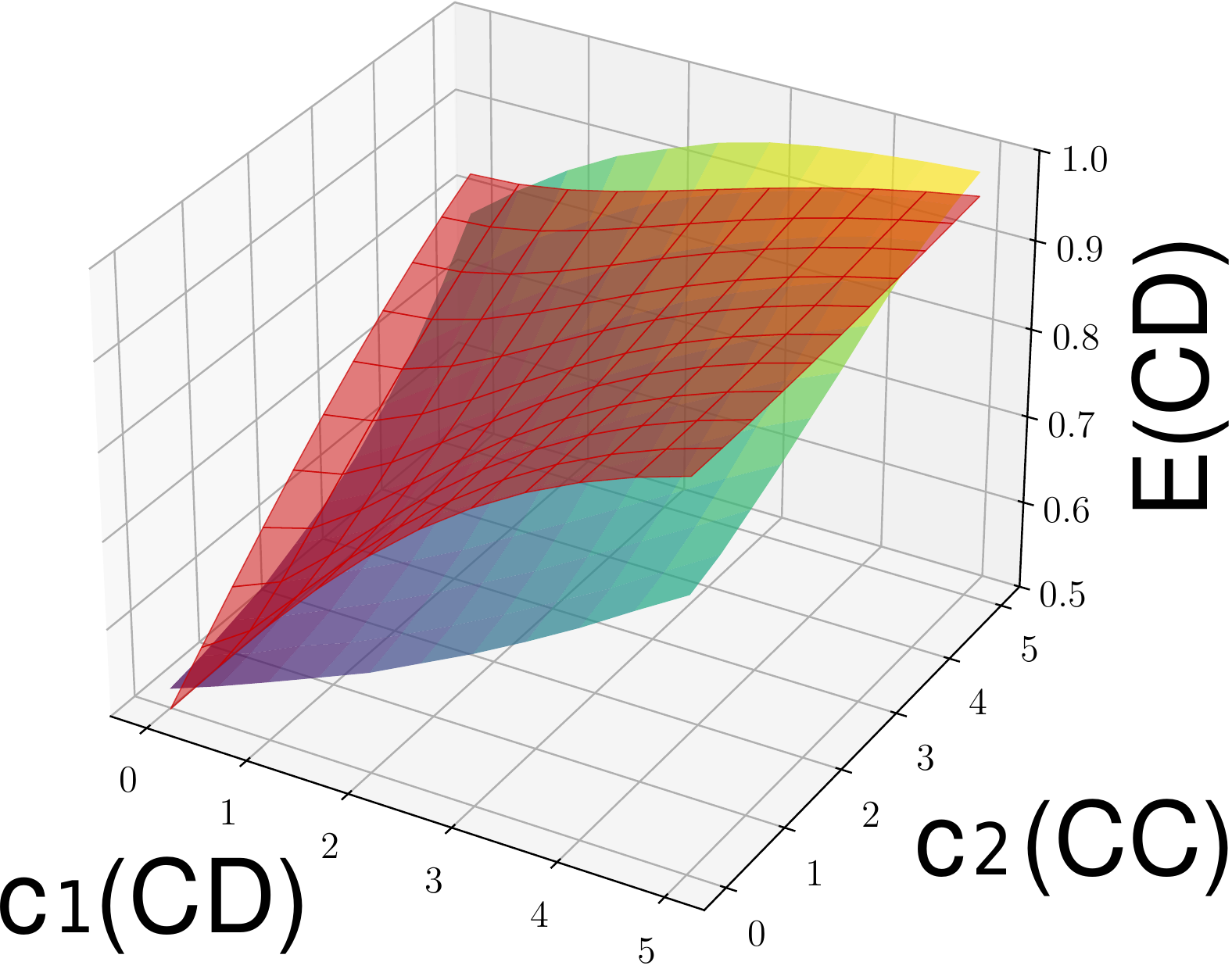}
    \\
    \includegraphics[width=.32\linewidth]{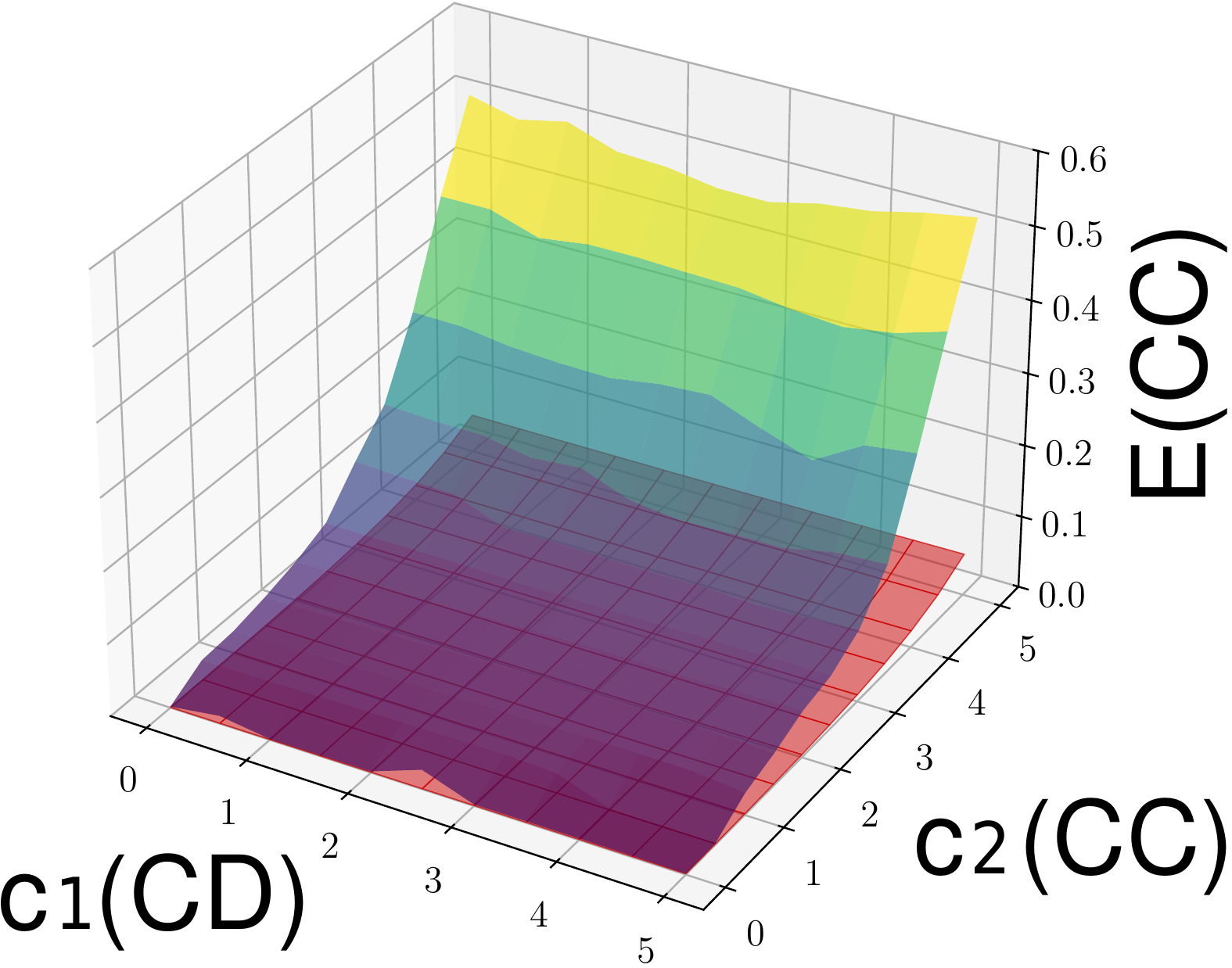}
    \includegraphics[width=.32\linewidth]{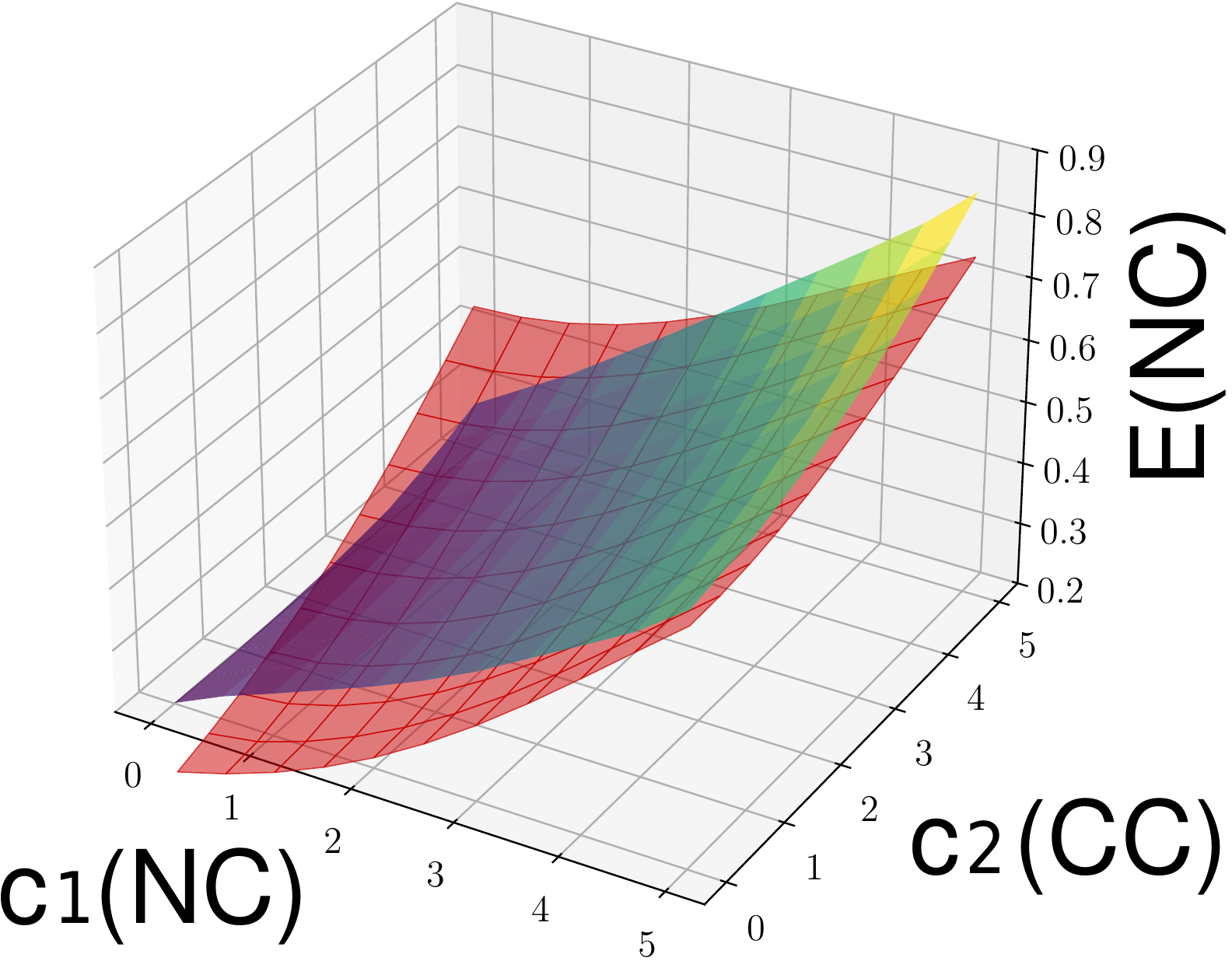}
    \includegraphics[width=.32\linewidth]{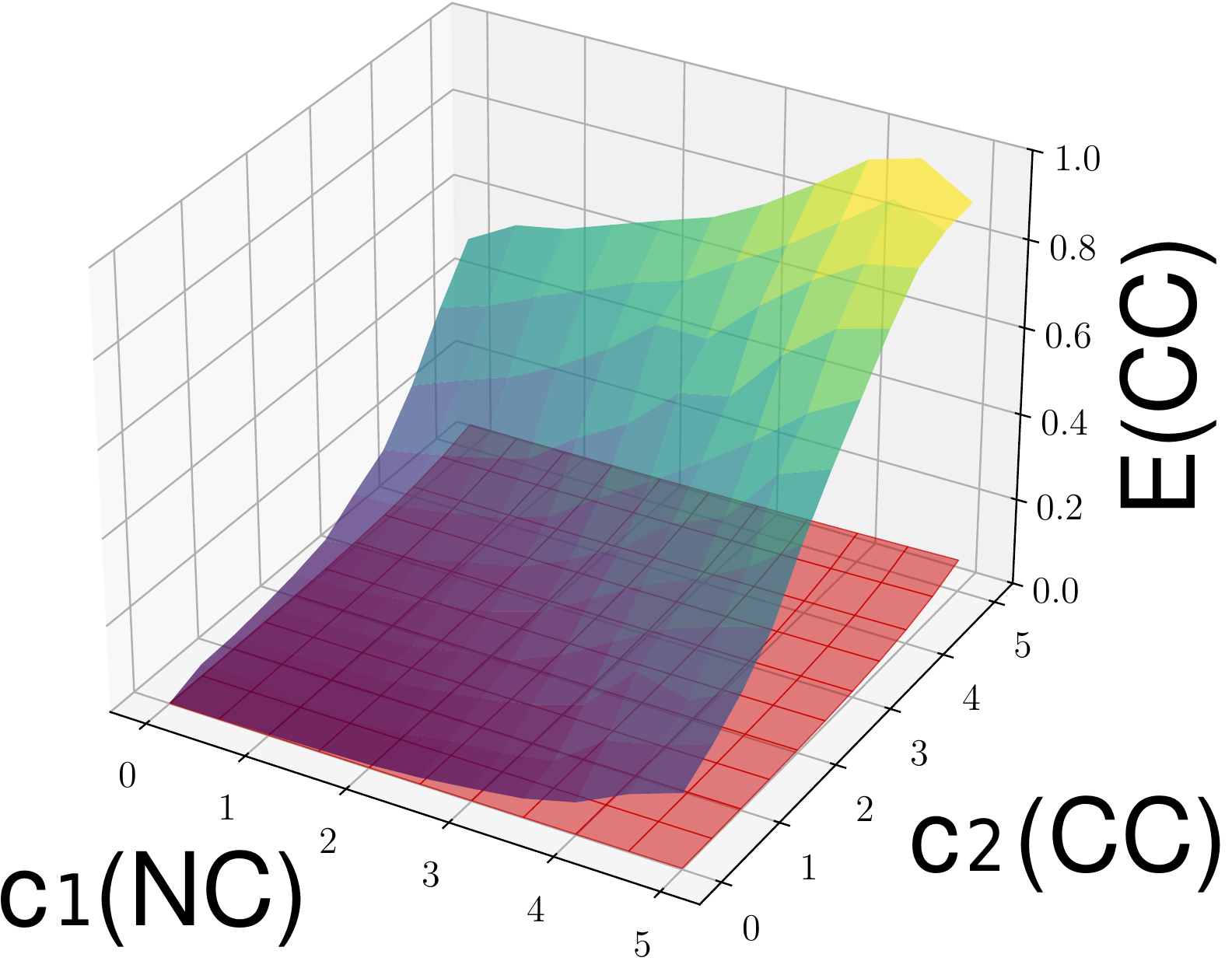}
    \end{minipage}
    \label{fig:cr_cr_pvcr_m1}
    }
    \\
    % 第二行：Qwen2.5coder_14b_Instruct模型
    \subfloat[Results of Qwen2.5coder\_14b\_Instruct model]{
    \begin{minipage}[t]{0.99\linewidth}
    \centering
    \includegraphics[width=0.32\linewidth]{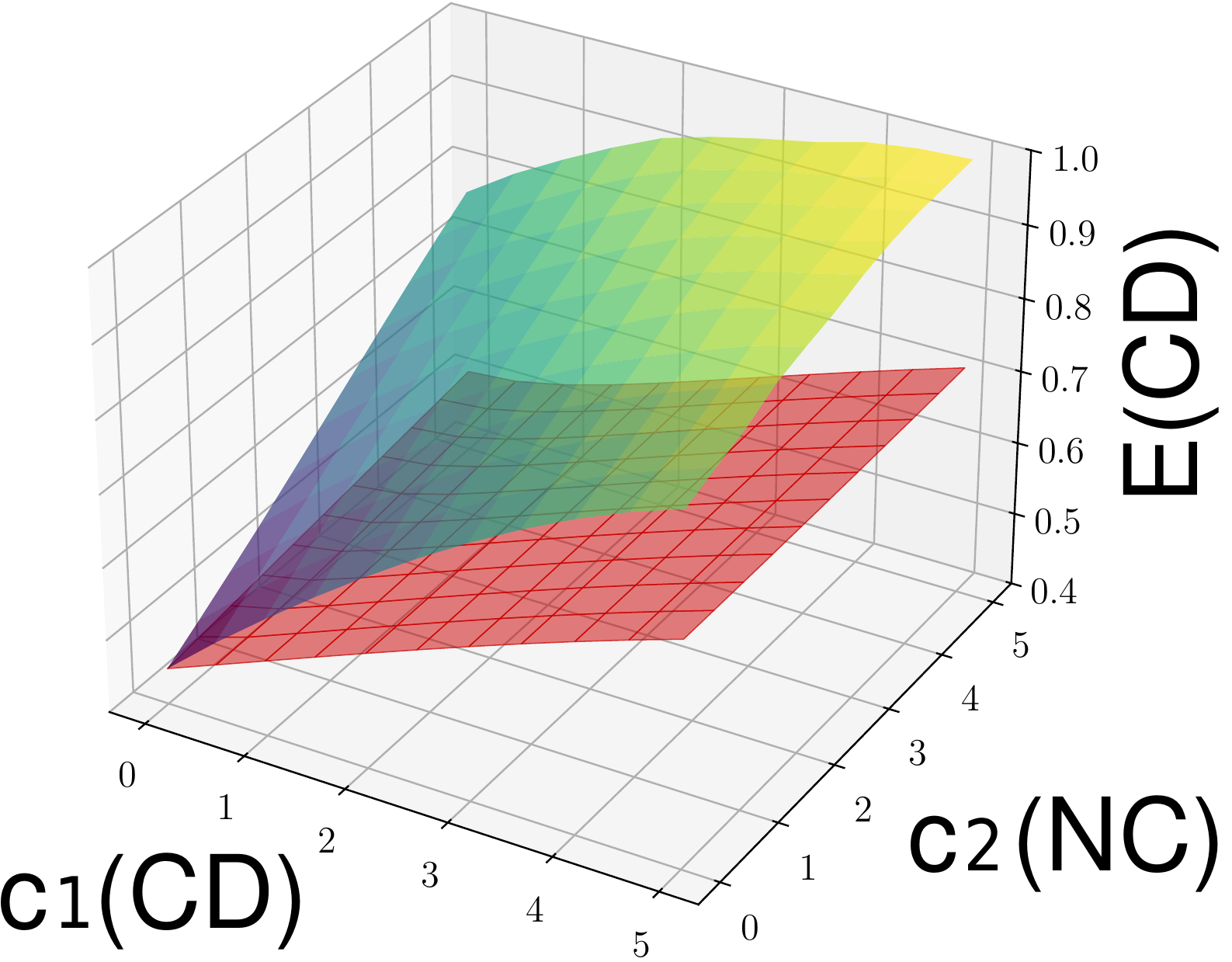}
    \includegraphics[width=0.32\linewidth]{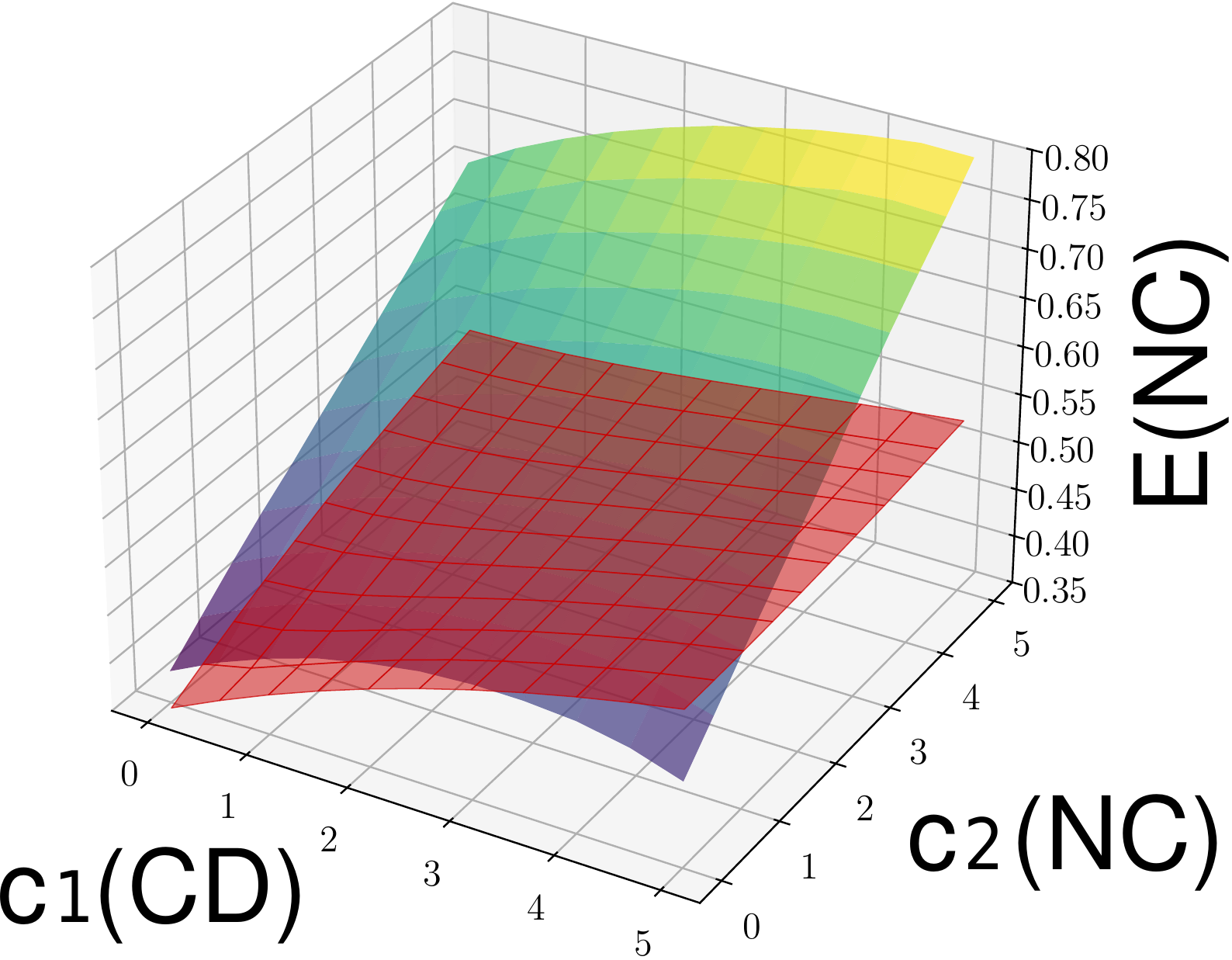}
    \includegraphics[width=0.32\linewidth]{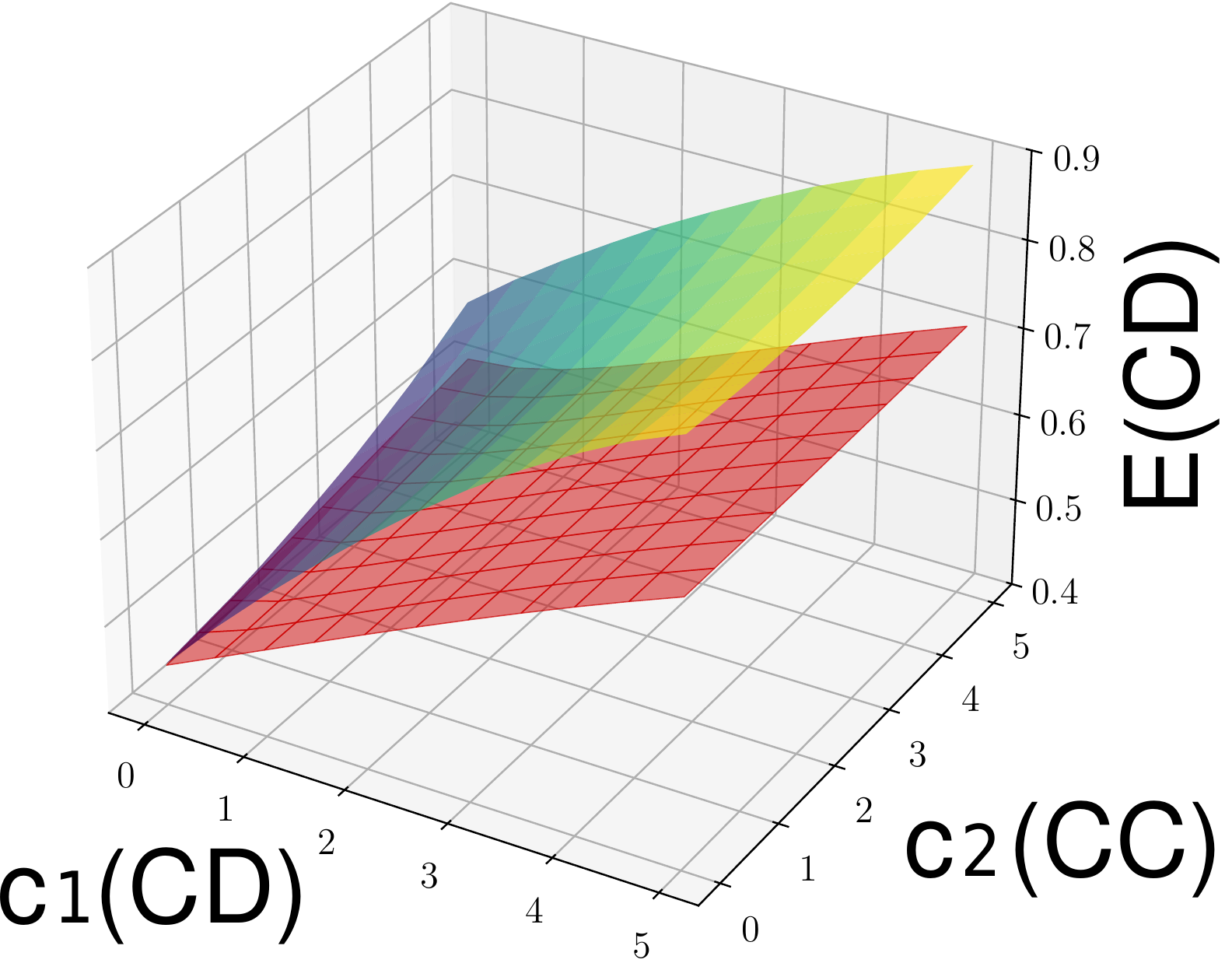}
    \\
    \includegraphics[width=0.32\linewidth]{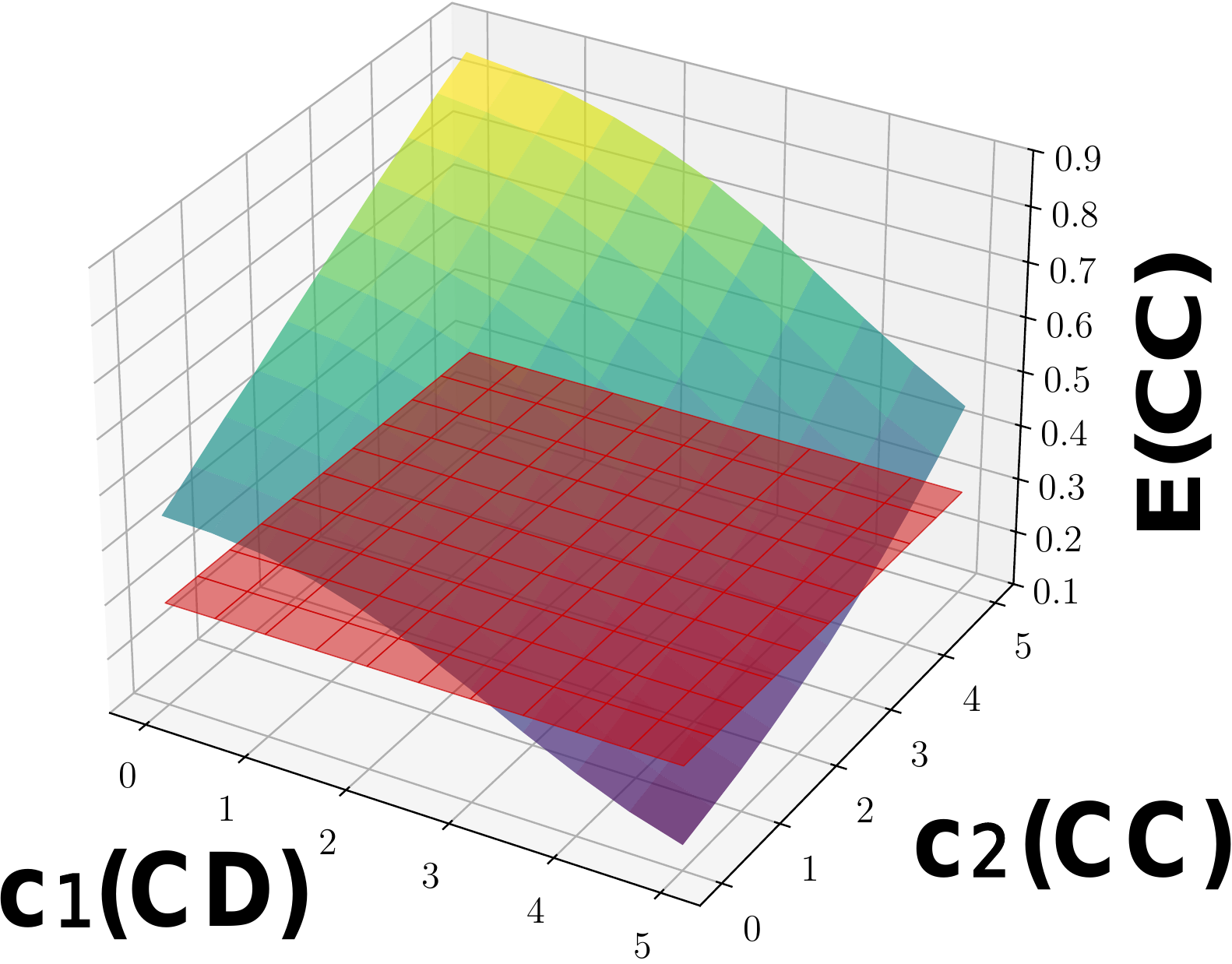}
    \includegraphics[width=0.32\linewidth]{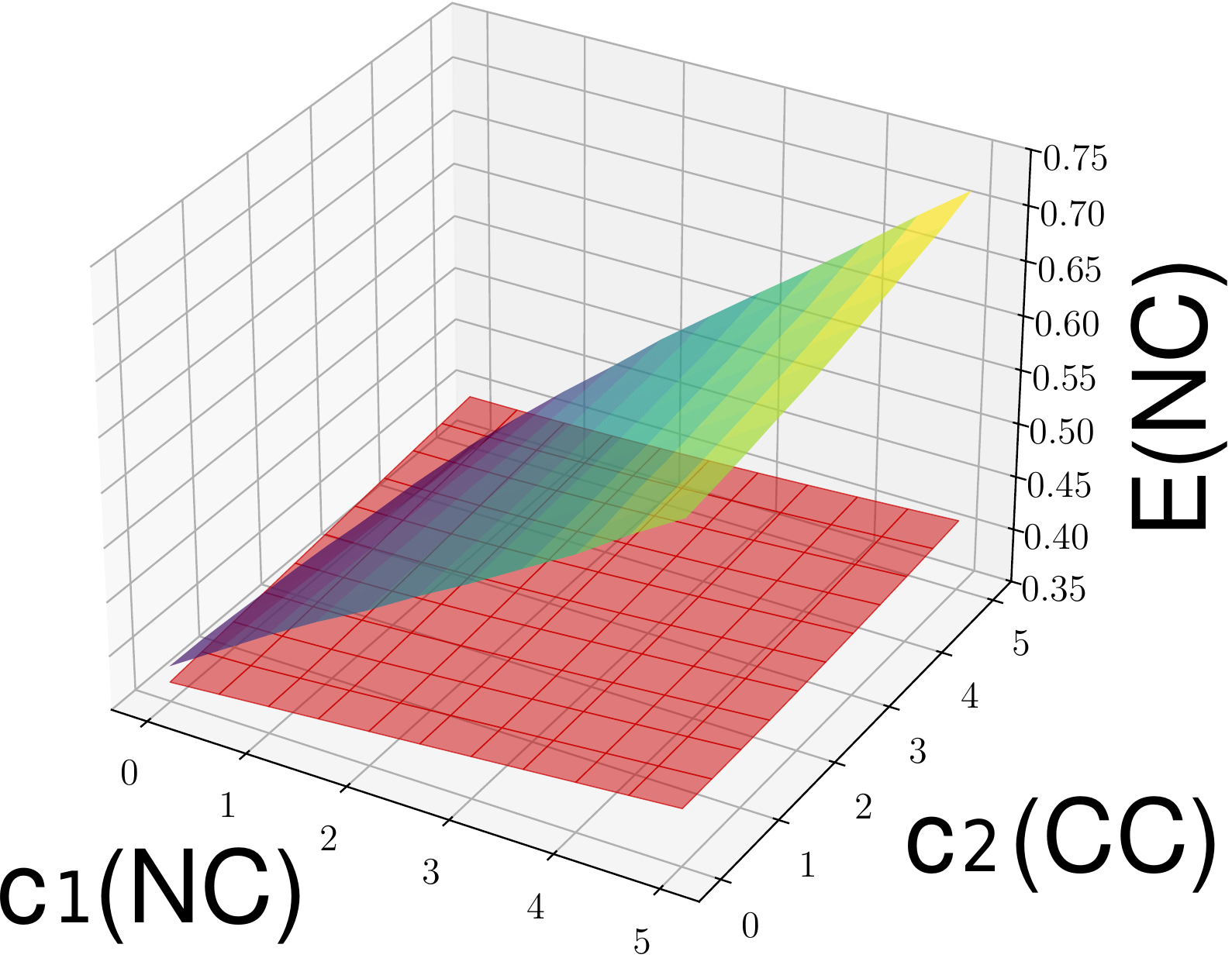}
    \includegraphics[width=0.32\linewidth]{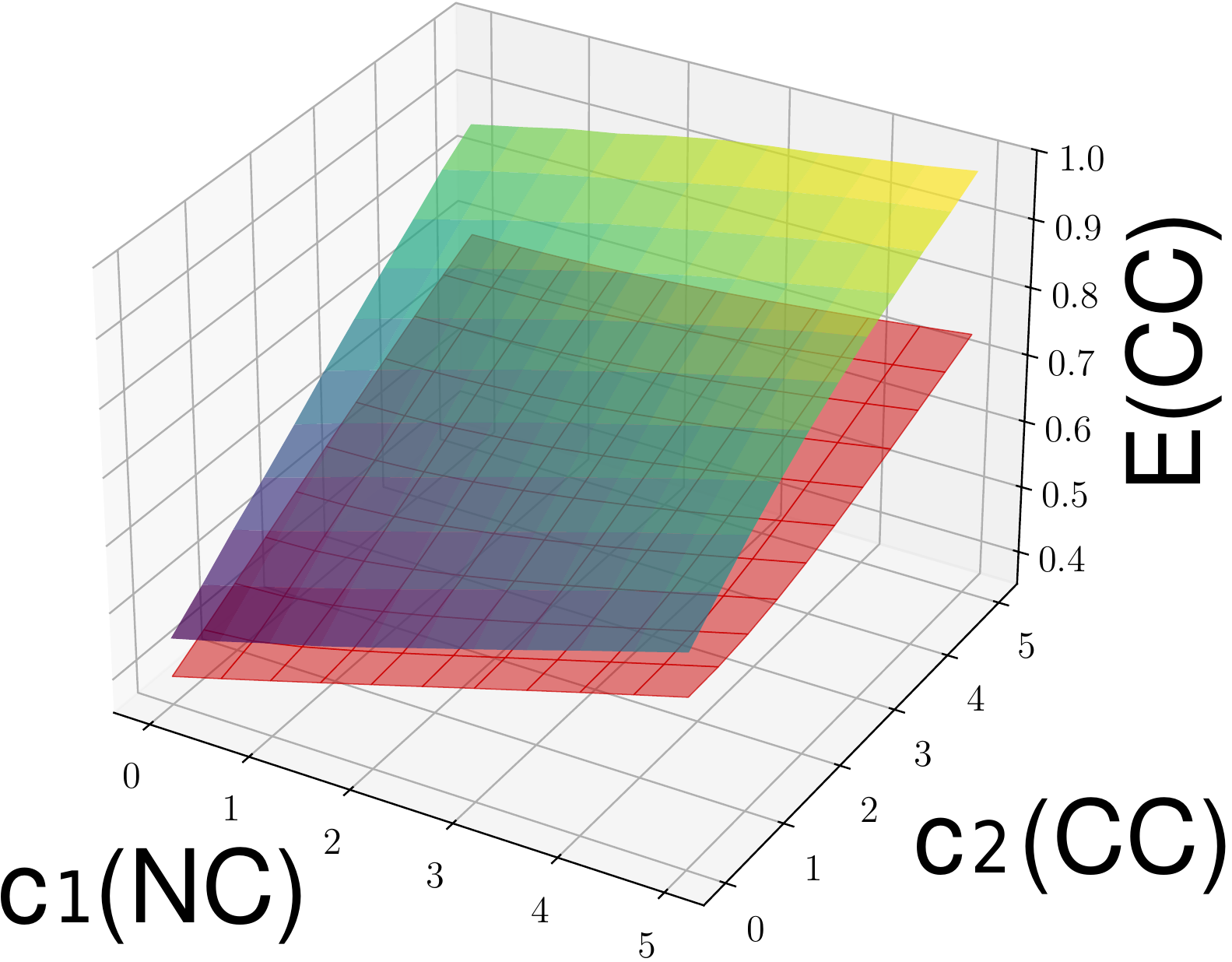}
    \end{minipage}
    \label{fig:cr_cr_pvcr_m3}
    }
    \\
    % 第三行：Codellama_13b_Instruct模型
    \subfloat[Results of Codellama\_13b\_Instruct model]{
    \begin{minipage}[t]{0.99\linewidth}
    \centering
    \includegraphics[width=0.32\linewidth]{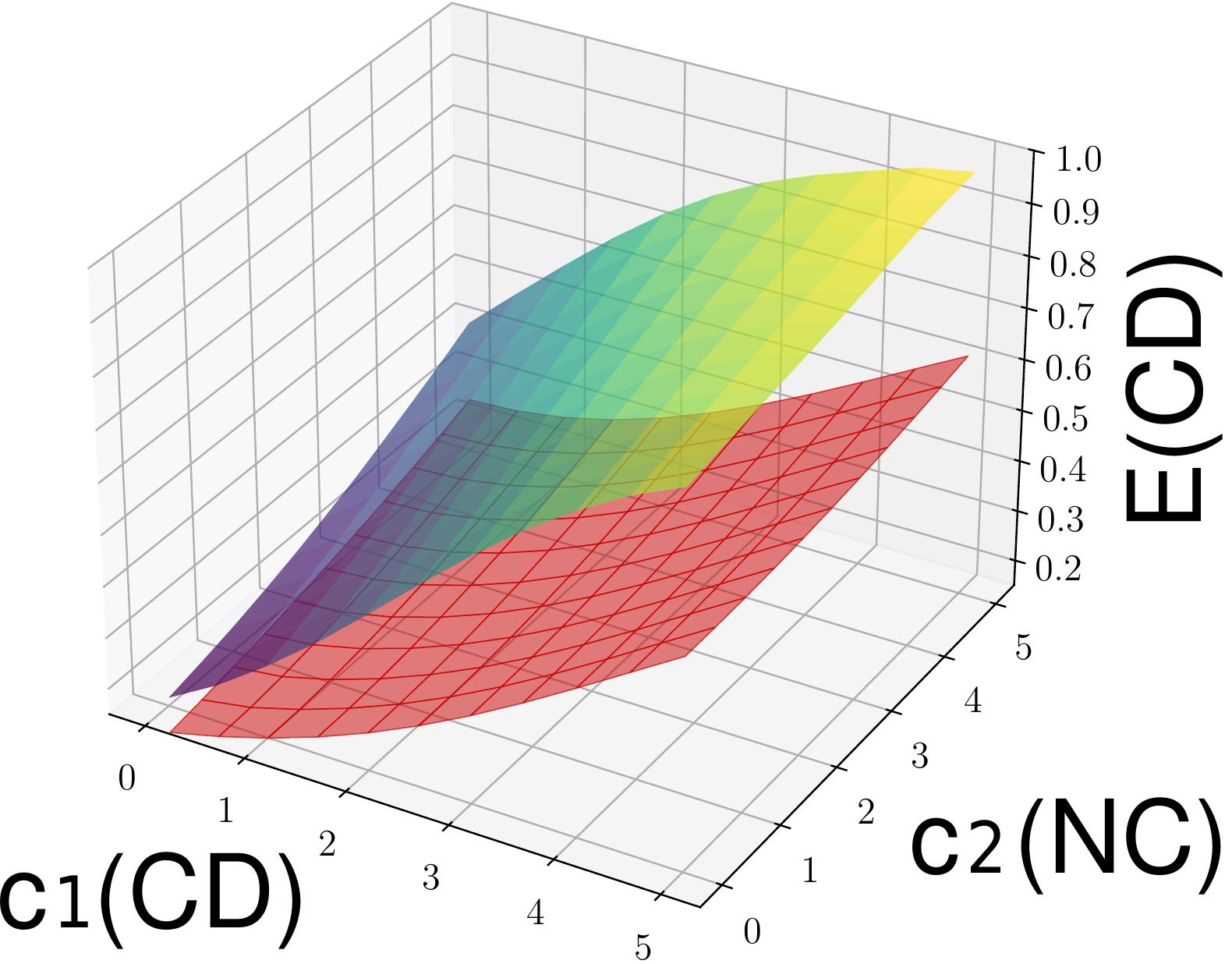}
    \includegraphics[width=0.32\linewidth]{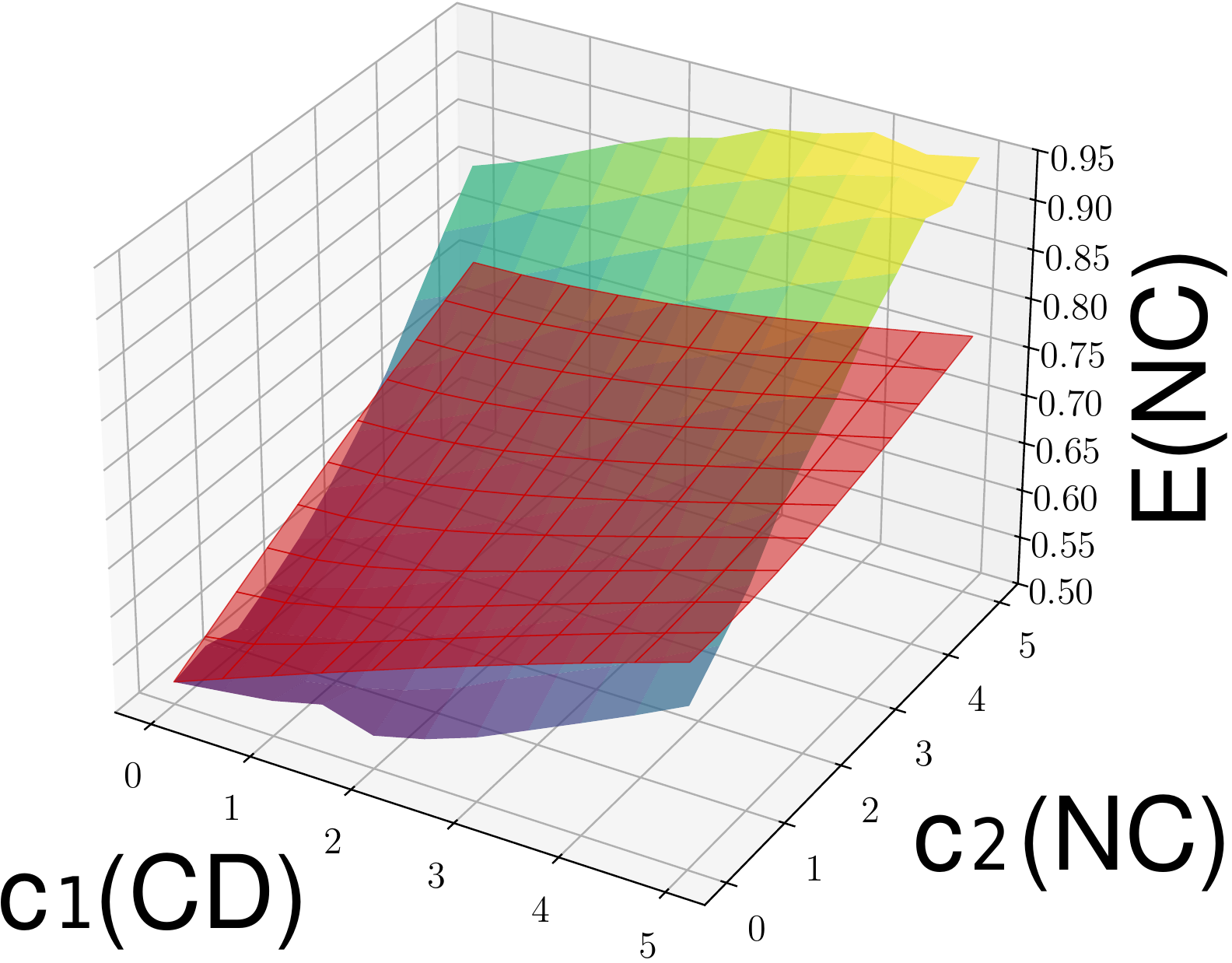}
    \includegraphics[width=0.32\linewidth]{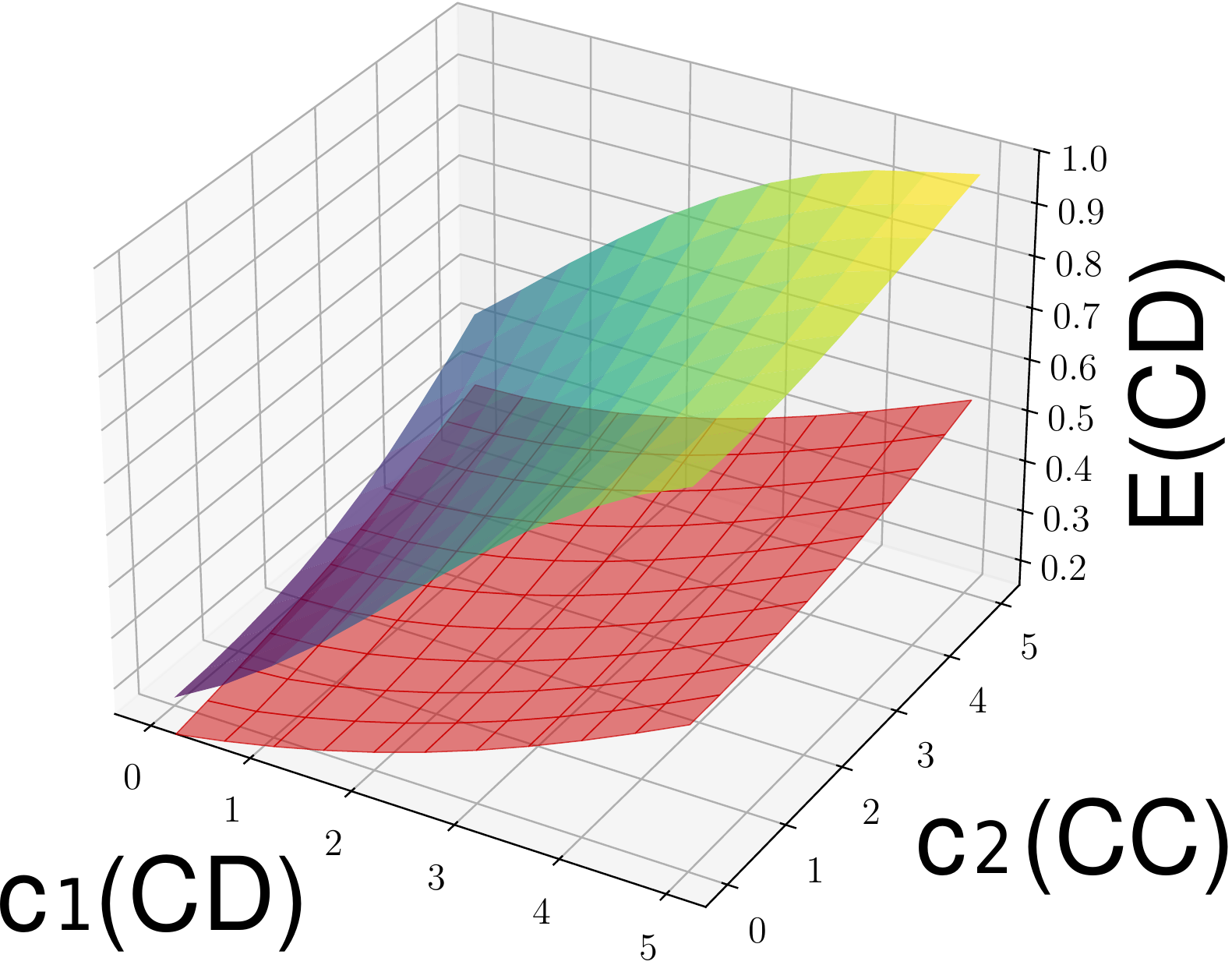}
    \\
    \includegraphics[width=0.32\linewidth]{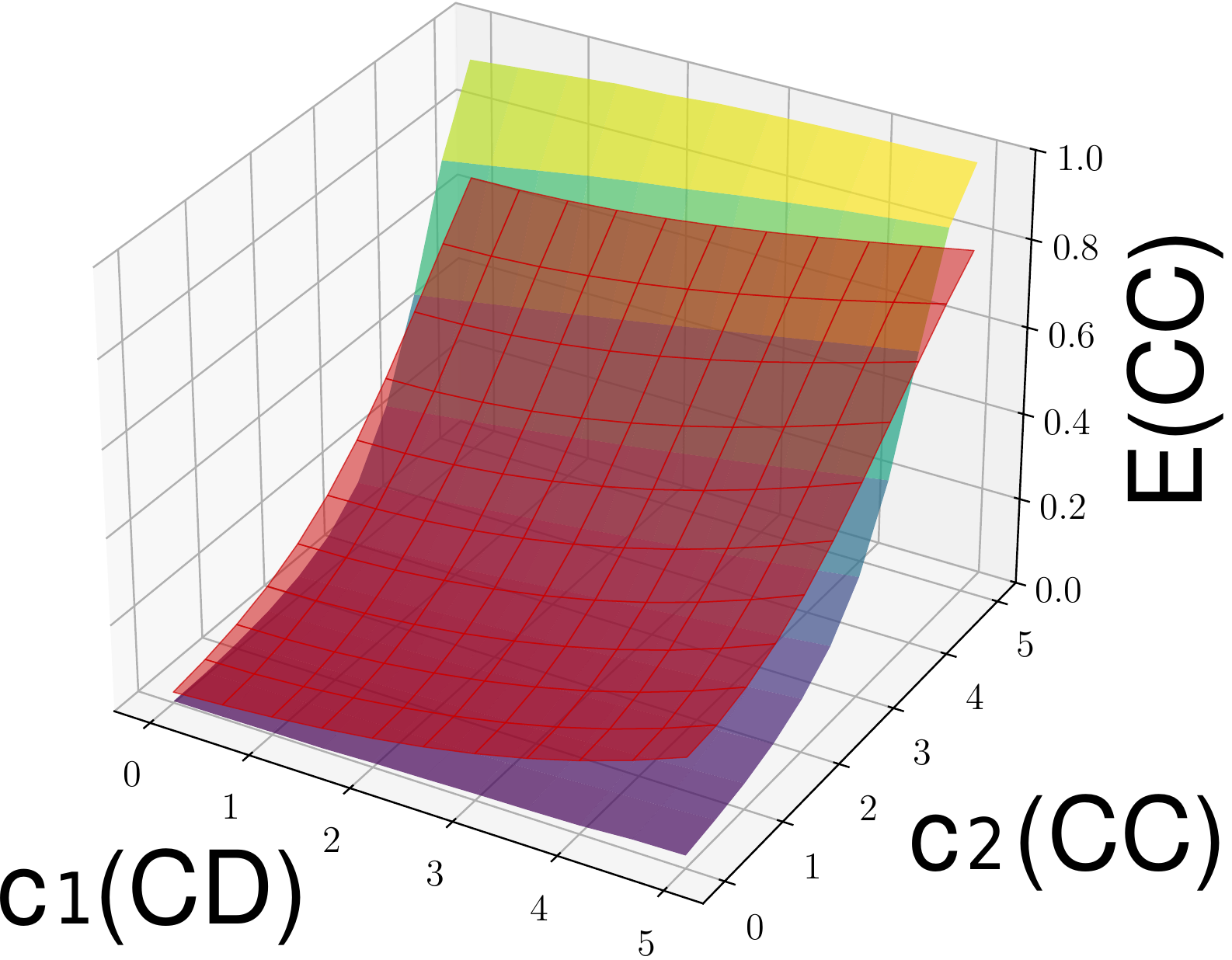}
    \includegraphics[width=0.32\linewidth]{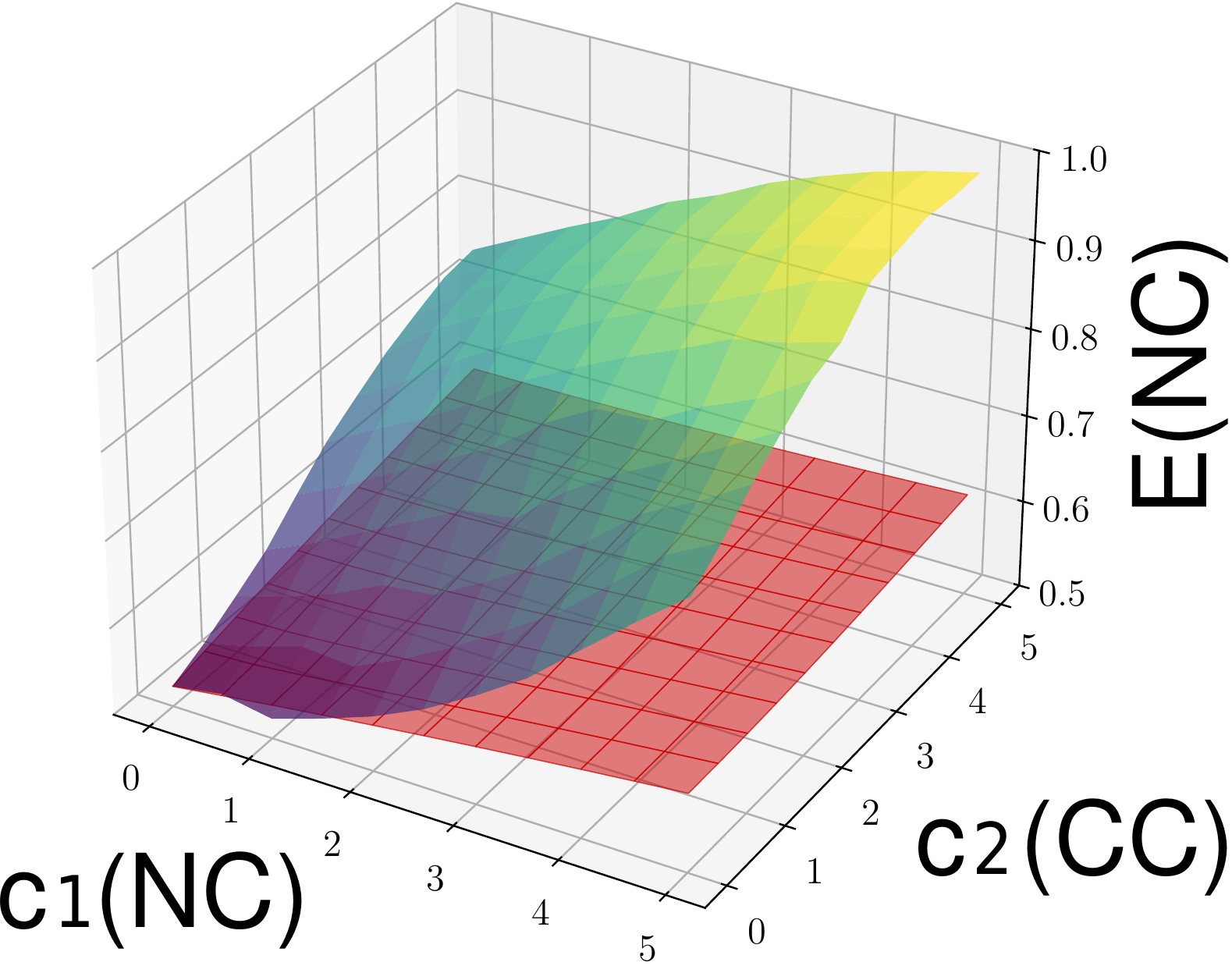}
    \includegraphics[width=0.32\linewidth]{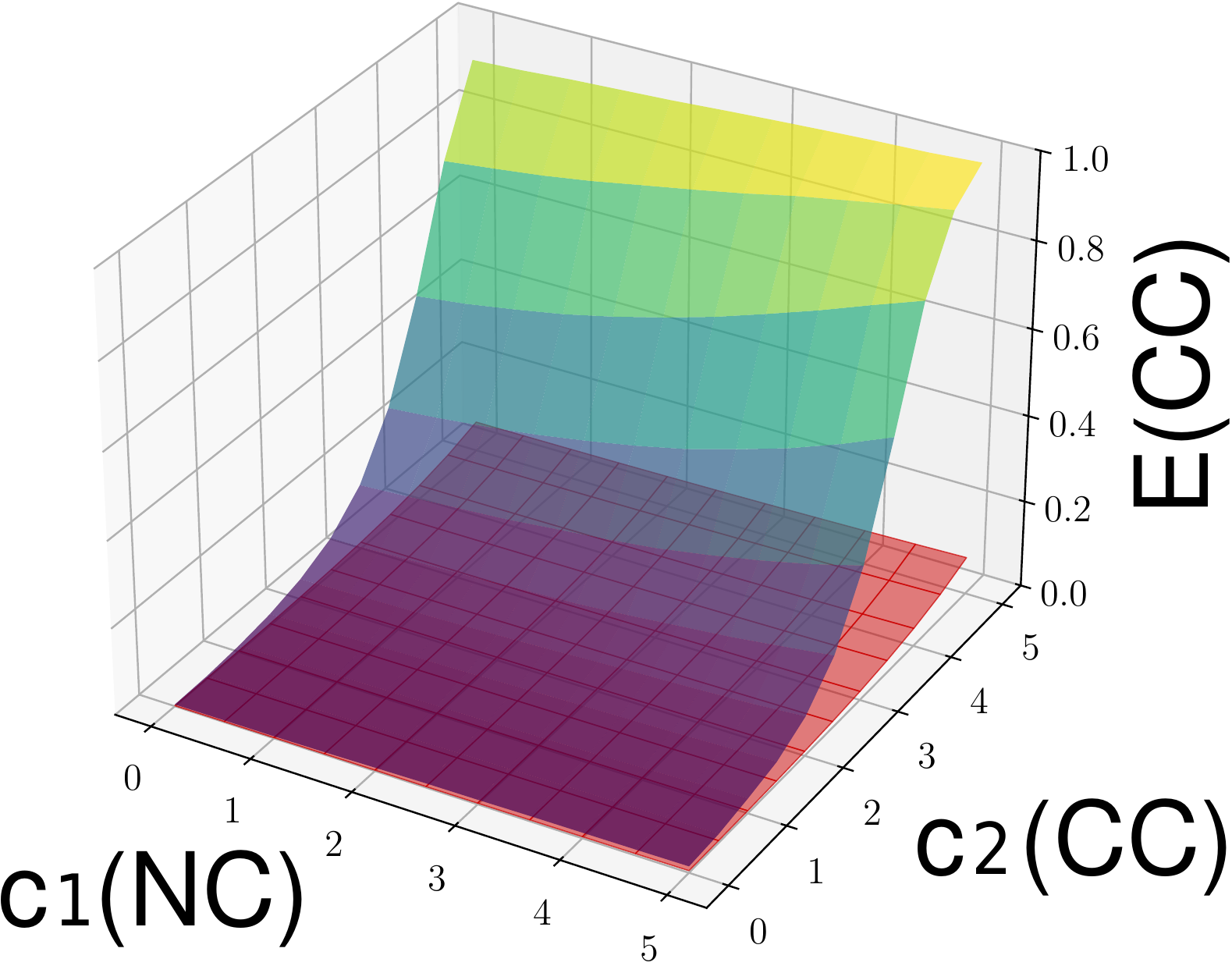}
    \end{minipage}
    \label{fig:cr_cr_pvcr_m4}
    }
    \caption{The surface plots illustrate how the code readability performance of the models varies with the coefficients of the injected steering vectors. The green surfaces represent the direct measurement results, and the red surfaces depict the theoretical bound given by Theorem~\ref{theorem1}.
    }\label{fig:cr_cr_pvcr}
\end{figure}

To demonstrate the performance on the code readability-related queries, we compute the expected probability that the model produces a positive answer. For instance, when a pair of the readability metrics, such as comment density and naming conventions, are injected, the model is queried with respect to the comment density and the expected probability that the answer aligns with high comment density is estimated. Figs.~\ref{fig:cr_cr_pvcr_m1},~\ref{fig:cr_cr_pvcr_m3}, and~\ref{fig:cr_cr_pvcr_m4} present the experimental results for the Deepseek\_R1\_14b, Qwen2.5coder\_14b\_Instruct, and Codellama\_13b\_Instruct models, respectively. The curves show how the expected probability of positive answers varies as the coefficients of the steering vectors increase. The fitted surfaces correspond to the lower bound predicted by Theorem~\ref{theorem1}. Tbl.~\ref{tab:statistically_significant_NLL} presents the injection layers of steering vectors corresponding to different readability metrics in each model, along with the statistical significance tests for model behavior changes.
\begin{table}[!t]
\centering
\small
\renewcommand{\arraystretch}{1.2}
\setlength{\tabcolsep}{2.25pt}
\caption{T-test Results and Steering Vector Injection Layers of Different Models}
\label{tab:t_test_results_with_layers}

\begin{tabular}{|c|c|c|c|c|}
\hline
\multirow{2}{*}{Model}
& \multirow{2}{*}{Metric}
& \multirow{2}{*}{Injection Layers}
& \multicolumn{2}{c|}{T-test Metrics} \\
\cline{4-5}
& & & T-statistic & $p$ \\
\hline

\multirow{3}{*}{DeepSeek-R1}
& CD & $-35$ to $-41$ & -5.73 & $2.85 \times 10^{-8}$ \\
\cline{2-5}
& NC & $-22$ to $-34$ & -5.58 & $6.13 \times 10^{-8}$ \\
\cline{2-5}
& CC & $-1$ to $-21$ & -5.71 & $3.16 \times 10^{-8}$ \\
\hline

\multirow{3}{*}{Qwen2.5-Coder}
& CD & $-15$ to $-43$ & -4.31 & $2.33 \times 10^{-5}$ \\
\cline{2-5}
& NC & $-13$ to $-19$ & -3.86 & $1.46 \times 10^{-4}$ \\
\cline{2-5}
& CC & $-20$ to $-35$ & -3.82 & $1.66 \times 10^{-4}$ \\
\hline

\multirow{3}{*}{CodeLlama}
& CD & $-20$ to $-24$ & -4.02 & $7.60 \times 10^{-5}$ \\
\cline{2-5}
& NC & $-30$ to $-31$ & -3.84 & $1.52 \times 10^{-4}$ \\
\cline{2-5}
& CC & $-25$ to $-29$ & -6.54 & $3.23 \times 10^{-10}$ \\
\hline

\end{tabular}
\label{tab:statistically_significant_NLL}
\end{table}

\zeng{The results indicate that by injecting the code readability steering vectors, all the tested models show a significant increase in the probability of providing positive answers to the readability-related queries, fully validating the positive regulatory effect of the readability steering vectors. Specifically, as the coefficients of the steering vectors increase, the models' recognition of the code readability metrics demonstrate a stable upward trend. The experimental data align with the theoretical lower bound in Theorem~\ref{theorem1}, indicating that the injection of steering vectors can effectively and consistently improve the code readability. Furthermore, differences in the sensitivity of various models to the steering vectors are observed, reflecting the influence of model architecture on the effectiveness of code readability regulation, thereby providing an empirical basis for subsequent optimization.}

\subsection{Code Correctness Measurement}\label{subsec:code_correctness_measurement_2sv}
\begin{figure}[!t]
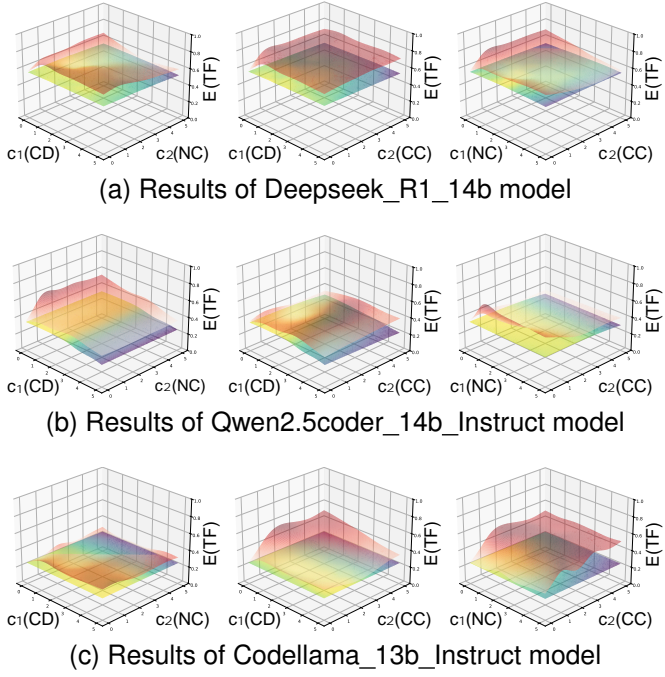

    \centering
    % 第一行：Deepseek_R1_14b模型
    \subfloat[Results of Deepseek\_R1\_14b model]{
    \begin{minipage}[t]{0.99\linewidth}
    \centering
    \includegraphics[width=.32\linewidth]{figs/CD_NC_TF_t2_Deepseek_R1_14b.pdf}
    \includegraphics[width=.32\linewidth]{figs/CD_CC_TF_t2_Deepseek_R1_14b.pdf}
    \includegraphics[width=.32\linewidth]{figs/NC_CC_TF_t2_Deepseek_R1_14b.pdf}
    \end{minipage}
    \label{fig:cr_cr_pvtf_m1}
    }
    \\
    % 第二行：Qwen2.5coder_14b_Instruct模型
    \subfloat[Results of Qwen2.5coder\_14b\_Instruct model]{
    \begin{minipage}[t]{0.99\linewidth}
    \centering
    \includegraphics[width=.32\linewidth]{figs/CD_NC_TF_t2_Qwen2.5coder_14b_Instruct.pdf}
    \includegraphics[width=.32\linewidth]{figs/CD_CC_TF_t2_Qwen2.5coder_14b_Instruct.pdf}
    \includegraphics[width=.32\linewidth]{figs/NC_CC_TF_t2_Qwen2.5coder_14b_Instruct.pdf}
    \end{minipage}
    \label{fig:cr_cr_pvtf_m3}
    }
    \\
    % 第三行：Codellama_13b_Instruct模型
    \subfloat[Results of Codellama\_13b\_Instruct model]{
    \begin{minipage}[t]{0.99\linewidth}
    \centering
    \includegraphics[width=0.32\linewidth]{figs/CD_NC_TF_t2_Codellama_13b_Instruct.pdf}
    \includegraphics[width=0.32\linewidth]{figs/CD_CC_TF_t2_Codellama_13b_Instruct.pdf}
    \includegraphics[width=0.32\linewidth]{figs/NC_CC_TF_t2_Codellama_13b_Instruct.pdf}
    \end{minipage}
    \label{fig:cr_cr_pvtf_m4}
    }
    \caption{The surface plots illustrate how the code correctness performance of the models varies with the coefficients of the injected steering vectors. The green surfaces represent the direct measurement results, and the red surfaces depict the theoretical bound given by Theorem~\ref{theorem2}.}\label{fig:cr_cr_pvtf}
\end{figure}

To show the performance on the code correctness-related queries, a procedure analogous to that for the code readability is adopted. For example, when the injected code readability metrics are comment density and naming conventions, the model is queried with respect to code correctness and the expected probability that the answer aligns with code correctness is computed. Figs.~\ref{fig:cr_cr_pvtf_m1},~\ref{fig:cr_cr_pvtf_m3}, and~\ref{fig:cr_cr_pvtf_m4} present the experimental results for the Deepseek\_R1\_14b, Qwen2.5coder\_14b\_Instruct, and Codellama\_13b\_Instruct models, respectively. The figures show how the expected probability of positive answers regarding the code correctness changes as the coefficients of the code readability steering vectors increase. The fitted surfaces correspond to the upper bound predicted by Theorem~\ref{theorem2}. {Details regarding how the parameters are determined can be found in Tables 1, 2, and 3 of the Appendix.}

\zeng{Similarly, the results demonstrate that,as the coefficients of the code readability steering vectors increase, the probability of all the tested models generating positive answers to the code correctness-related queries shows a declining trend, which aligns with the upper bound prescribed by Theorem~\ref{theorem2}. Specifically, when steering vectors corresponding to any two code readability metrics are injected, the  correctness performance gradually decreases as the coefficients grow, but the extent of this decline is constrained by the theoretical upper bound. This finding confirms that through controlled steering vector injection, it is possible to improve the code readability while limiting the loss in the code correctness within the theoretical bound, providing a crucial basis for the parameter tuning in practical applications.}

\subsection{Trade-off between the Readability and Correctness}\label{subsec:code_quality_measurement_fsv}
\begin{figure}[!t]
    \centering
    % 第一行：Deepseek_R1_14b模型
    \subfloat[Results of Deepseek\_R1\_14b model]{
    \begin{minipage}[t]{0.99\linewidth}
    \centering
    \includegraphics[width=.34\linewidth]{figs/CD_NC_CC_CD_Deepseek_R1_14b.pdf}
    \includegraphics[width=.32\linewidth]{figs/CD_NC_CC_NC_Deepseek_R1_14b.pdf}
    \includegraphics[width=.32\linewidth]{figs/CD_NC_CC_CC_Deepseek_R1_14b.pdf}
    \end{minipage}
    \label{fig:cr_cr_cr_pv_m1}
    }
    \\
    % 第二行：Qwen2.5coder_14b_Instruct模型
    \subfloat[Results of Qwen2.5coder\_14b\_Instruct model]{
    \begin{minipage}[t]{0.99\linewidth}
    \centering
    \includegraphics[width=.32\linewidth]{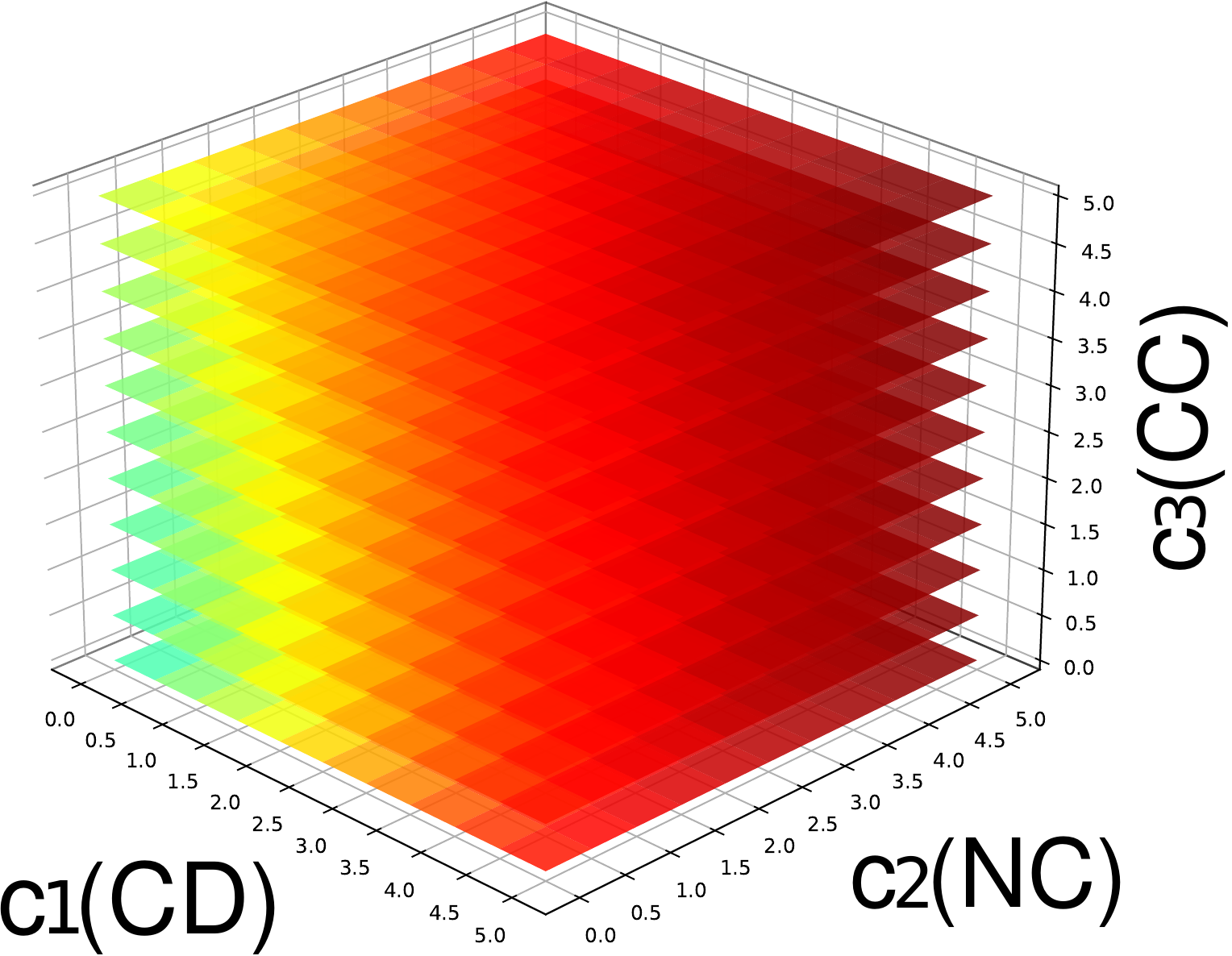}
    \includegraphics[width=.32\linewidth]{figs/CD_NC_CC_NC_Qwen2.5coder_14b_Instruct.pdf}
    \includegraphics[width=.32\linewidth]{figs/CD_NC_CC_CC_Qwen2.5coder_14b_Instruct.pdf}
    \end{minipage}
    \label{fig:cr_cr_cr_pv_m3}
    }
    \\
    % 第三行：Codellama_13b_Instruct模型
    \subfloat[Results of Codellama\_13b\_Instruct model]{
    \begin{minipage}[t]{0.99\linewidth}
    \centering
    \includegraphics[width=0.32\linewidth]{figs/CD_NC_CC_CD_Codellama_13b_Instruct.pdf}
    \includegraphics[width=0.32\linewidth]{figs/CD_NC_CC_NC_Codellama_13b_Instruct.pdf}
    \includegraphics[width=0.32\linewidth]{figs/CD_NC_CC_CC_Codellama_13b_Instruct.pdf}
    \end{minipage}
    \label{fig:cr_cr_cr_pv_m4}
    }
    \caption{The 3D stacked heatmaps illustrate the probability~(blue for $0$, red for $1$) of models, injected with steering vectors corresponding to all three code readability metrics, providing positive answers to various queries under different coefficient values. The queries correspond to comment density, naming conventions, and cyclomatic complexity, respectively, from the first to the third column.}\label{fig:cr_cr_cr_pv}
\end{figure}

\begin{figure}[!t]
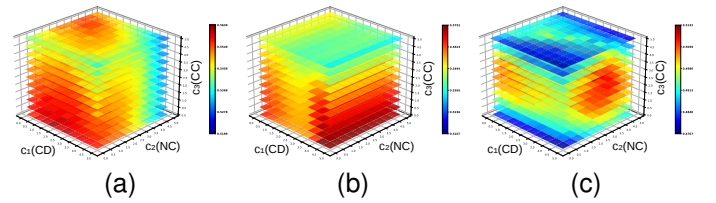

    \centering
    \subfloat[]{
    \includegraphics[width=.32\linewidth]{figs/CD_NC_CC_TF_Deepseek_R1_14b.pdf}
    \label{fig:cd_nc_cc_tf_m1}
    }
    \subfloat[]{
    \includegraphics[width=.32\linewidth]{figs/CD_NC_CC_TF_Qwen2.5coder_14b_Instruct.pdf}
    \label{fig:cd_nc_cc_tf_m3}
    }
    \subfloat[]{
    \includegraphics[width=.32\linewidth]{figs/CD_NC_CC_TF_Codellama_13b_Instruct.pdf}
    \label{fig:cd_nc_cc_tf_m4}
    }
    \caption{The 3D stacked heatmaps illustrate the probability~(blue for $0$, red for $1$) of models, injected with steering vectors corresponding to all three code readability metrics, providing positive answers to queries related to code correctness under varying coefficients.}\label{fig:cd_nc_cc_tf}
\end{figure}

{We analyze the impact on the models' performance after the steering vector injection in terms of the three readability metrics.} \yu{\autoref{fig:cr_cr_cr_pv} shows the probability of Deepseek\_R1\_14b, Qwen2.5coder\_14b\_Instruct, and Codellama\_13b\_Instruct, which provides positive answers to varied code readability-related queries after injecting steering vectors in terms of the code readability metrics} across different coefficient values. \yu{We also provide the corresponding probabilities of the three models that generate positive answers to code correctness queries in~\autoref{fig:cd_nc_cc_tf}}. It can be observed that for each code readability metric, its corresponding steering vector plays a dominant role in improving the model's performance . Furthermore, as the coefficients of the code readability steering vectors increase, the model's performance on code correctness exhibits a declining trend.

\zeng{We can systematically optimize the trade-off between the code readability and correctness by collaboratively leveraging Theorems~\ref{theorem1} and~\ref{theorem2}. Theorem~\ref{theorem1} provides a clear lower-bound guarantee for improving the readability, ensuring that after injecting steering vectors, the model's probability of positive answers to the readability-related queries increases at least at the theoretically predicted rate, thereby reliably improving the  readability metrics. Theorem~\ref{theorem2} constrains potential negative impacts by setting an upper bound for the decline in the code correctness, preventing excessive degradation of correctness. Specifically, by adjusting the coefficients of steering vectors corresponding to different  readability metrics, we can achieve a fine-grained balance between the degree of the readability improvement and the tolerable level of the correctness loss, tailored to specific application scenarios. For instance, the experimental results show that for the Deepseek\_R1\_14b model, with the coefficient combination ${\rm c}_1 ({\rm CD}) = 1.0$, ${\rm c}_2 ({\rm NC}) = 4.0$, and ${\rm c}_3 ({\rm CC}) = 5.0$, the  readability is significantly improved while the correctness decreases by only $0.79\%$ — a trade-off acceptable in most practical applications. Similarly, the Qwen2.5coder\_14b\_Instruct model achieves readability optimization with ${\rm c}_1 ({\rm CD}) = 3.5$, ${\rm c}_2 ({\rm NC}) = 5.0$, and ${\rm c}_3 ({\rm CC}) = 3.0$, keeping the correctness loss within $0.95\%$. The Codellama\_13b\_Instruct model exhibits a mere $0.28\%$ the correctness loss with ${\rm c}_1 ({\rm CD}) = 5.0$, ${\rm c}_2 ({\rm NC}) = 4.5$, and ${\rm c}_3 ({\rm CC}) = 5.0$. These results validate the effectiveness of the steering strategy based on the theoretical bounds, enabling MRepE to achieve a controllable balance between the readability and correctness tailored to different model characteristics and application requirements.}

\section{Conclusion and Future Work}\label{sec:conclusion}
This work addresses the control problem of multiple readability metrics in LLM-generated codes by defining a joint principal component analysis problem under multi-dimensional constraints, proposing a corresponding solution algorithm~(MOC-JPCA), and developing a multitask steering representation engineering method~(MRepE) to coordinately control multiple readability metrics. 
By using MOC-JPCA to jointly extract steering vectors for different code readability metrics and applying them in the MRepE, we achieve coordinated control over these metrics and provide theoretical analysis of the trade-off between the code readability and correctness. The experimental results further show that the extracted steering vectors can consistently improve the model's responses to readability-related queries, while the degradation in code correctness can be limited by properly selecting the steering coefficients. This indicates that MRepE provides not only an effective method for multi-metric readability control, but also a controllable mechanism for balancing human-oriented code quality and functional reliability. We could refine the code generation according to users' demands. Although we investigate the MRepE framework over three readability metrics, MRepE can be generalized to more code quality control tasks under similar theoretical bounds on its performance of the code readability and correctness. In future work, we plan to extend MRepE to more fine-grained code quality attributes, such as maintainability, modularity, and documentation style, and evaluate its effectiveness in larger-scale software engineering scenarios. Future work will focus on practical applications of the MRepE and investigate its synergy with different aspects of LLM-generated code quality.

\bibliographystyle{IEEEtran}
\bibliography{ref}

@article{zou2023representation,
  title={Representation engineering: a top-down approach to ai transparency},
  author={Zou, Andy and Phan, Long and Chen, Sarah and Campbell, James and Guo, Phillip and Ren, Richard and Pan, Alexander and Yin, Xuwang and Mazeika, Mantas and Dombrowski, Ann-Kathrin and others},
  journal={ArXiv Preprint arXiv:2310.01405},
  year={2023}
}

@inproceedings{wolf2025tradeoffs,
  title={Tradeoffs Between Alignment and Helpfulness in Language Models with Steering Methods},
  author={Wolf, Yotam and Wies, Noam and Shteyman, Dorin and Rothberg, Binyamin and Levine, Yoav and Shashua, Amnon},
  booktitle={ICLR 2025 Workshop on Foundation Models in the Wild},
  year={2025}
}

@article{jiang2024survey,
  title={A survey on large language models for code generation},
  author={Jiang, Juyong and Wang, Fan and Shen, Jiasi and Kim, Sungju and Kim, Sunghun},
  journal={arXiv preprint arXiv:2406.00515},
  year={2024}
}

@article{roziere2023code,
  title={Code llama: Open foundation models for code},
  author={Roziere, Baptiste and Gehring, Jonas and Gloeckle, Fabian and Sootla, Sten and Gat, Itai and Tan, Xiaoqing Ellen and Adi, Yossi and Liu, Jingyu and Sauvestre, Romain and Remez, Tal and others},
  journal={arXiv preprint arXiv:2308.12950},
  year={2023}
}

@article{bui2025correctness,
  title={Correctness Assessment of Code Generated by Large Language Models Using Internal Representations},
  author={Bui, Tuan-Dung and Vu, Thanh Trong and Nguyen, Thu-Trang and Nguyen, Son and Vo, Hieu Dinh},
  journal={arXiv preprint arXiv:2501.12934},
  year={2025}
}

@article{buse2009learning,
  title={Learning a metric for code readability},
  author={Buse, Raymond PL and Weimer, Westley R},
  journal={IEEE Transactions on software engineering},
  volume={36},
  number={4},
  pages={546--558},
  year={2009},
  publisher={IEEE}
}

@inproceedings{posnett2011simpler,
  title={A simpler model of software readability},
  author={Posnett, Daryl and Hindle, Abram and Devanbu, Premkumar},
  booktitle={Proceedings of the 8th working conference on mining software repositories},
  pages={73--82},
  year={2011}
}

@inproceedings{scalabrino2017automatically,
  title={Automatically assessing code understandability: How far are we?},
  author={Scalabrino, Simone and Bavota, Gabriele and Vendome, Christopher and Linares-V{\'a}squez, Mario and Poshyvanyk, Denys and Oliveto, Rocco},
  booktitle={2017 32nd IEEE/ACM International Conference on Automated Software Engineering (ASE)},
  pages={417--427},
  year={2017},
  organization={IEEE}
}

@article{guo2025deepseek,
  title={Deepseek-r1: Incentivizing reasoning capability in llms via reinforcement learning},
  author={Guo, Daya and Yang, Dejian and Zhang, Haowei and Song, Junxiao and Zhang, Ruoyu and Xu, Runxin and Zhu, Qihao and Ma, Shirong and Wang, Peiyi and Bi, Xiao and others},
  journal={arXiv preprint arXiv:2501.12948},
  year={2025}
}

@article{hui2024qwen2,
  title={Qwen2. 5-coder technical report},
  author={Hui, Binyuan and Yang, Jian and Cui, Zeyu and Yang, Jiaxi and Liu, Dayiheng and Zhang, Lei and Liu, Tianyu and Zhang, Jiajun and Yu, Bowen and Lu, Keming and others},
  journal={arXiv preprint arXiv:2409.12186},
  year={2024}
}

@article{luo2024badam,
  title={BAdam: A memory efficient full parameter optimization method for large language models},
  author={Luo, Qijun and Yu, Hengxu and Li, Xiao},
  journal={Advances in Neural Information Processing Systems},
  volume={37},
  pages={24926--24958},
  year={2024}
}

@inproceedings{qin2024federated,
  title={Federated Full-Parameter Tuning of Billion-Sized Language Models with Communication Cost under 18 Kilobytes},
  author={Qin, Zhen and Chen, Daoyuan and Qian, Bingchen and Ding, Bolin and Li, Yaliang and Deng, Shuiguang},
  booktitle={International Conference on Machine Learning},
  pages={41473--41497},
  year={2024},
  organization={PMLR}
}

@inproceedings{lv2024full,
  title={Full Parameter Fine-tuning for Large Language Models with Limited Resources},
  author={Lv, Kai and Yang, Yuqing and Liu, Tengxiao and Guo, Qipeng and Qiu, Xipeng},
  booktitle={Proceedings of the 62nd Annual Meeting of the Association for Computational Linguistics (Volume 1: Long Papers)},
  pages={8187--8198},
  year={2024}
}

@article{dettmers2023qlora,
  title={Qlora: Efficient finetuning of quantized llms},
  author={Dettmers, Tim and Pagnoni, Artidoro and Holtzman, Ari and Zettlemoyer, Luke},
  journal={Advances in neural information processing systems},
  volume={36},
  pages={10088--10115},
  year={2023}
}

@inproceedings{zhang2023adaptive,
  title={Adaptive Budget Allocation for Parameter-Efficient Fine-Tuning},
  author={Zhang, Qingru and Chen, Minshuo and Bukharin, Alexander and He, Pengcheng and Cheng, Yu and Chen, Weizhu and Zhao, Tuo},
  booktitle={International Conference on Learning Representations},
  year={2023},
  organization={Openreview}
}

@inproceedings{zhang2024llama,
  title={LLaMA-adapter: Efficient fine-tuning of large language models with zero-initialized attention},
  author={Zhang, Renrui and Han, Jiaming and Liu, Chris and Zhou, Aojun and Lu, Pan and Qiao, Yu and Li, Hongsheng and Gao, Peng},
  booktitle={The Twelfth International Conference on Learning Representations},
  year={2024}
}

@article{ouyang2022training,
  title={Training language models to follow instructions with human feedback},
  author={Ouyang, Long and Wu, Jeffrey and Jiang, Xu and Almeida, Diogo and Wainwright, Carroll and Mishkin, Pamela and Zhang, Chong and Agarwal, Sandhini and Slama, Katarina and Ray, Alex and others},
  journal={Advances in neural information processing systems},
  volume={35},
  pages={27730--27744},
  year={2022}
}

@article{yuan2023rrhf,
  title={Rrhf: Rank responses to align language models with human feedback},
  author={Yuan, Hongyi and Yuan, Zheng and Tan, Chuanqi and Wang, Wei and Huang, Songfang and Huang, Fei},
  journal={Advances in Neural Information Processing Systems},
  volume={36},
  pages={10935--10950},
  year={2023}
}

@inproceedings{hong2024orpo,
  title={ORPO: Monolithic Preference Optimization without Reference Model},
  author={Hong, Jiwoo and Lee, Noah and Thorne, James},
  booktitle={Proceedings of the 2024 Conference on Empirical Methods in Natural Language Processing},
  pages={11170--11189},
  year={2024}
}

@inproceedings{wang2023review,
  title={A review on code generation with llms: Application and evaluation},
  author={Wang, Jianxun and Chen, Yixiang},
  booktitle={2023 IEEE International Conference on Medical Artificial Intelligence (MedAI)},
  pages={284--289},
  year={2023},
  organization={IEEE}
}

@article{van2024extending,
  title={Extending activation steering to broad skills and multiple behaviours},
  author={van der Weij, Teun and Poesio, Massimo and Schoots, Nandi},
  journal={arXiv preprint arXiv:2403.05767},
  year={2024}
}

@phdthesis{stiefel1935richtungsfelder,
  title={Richtungsfelder und Fernparallelismus in n-dimensionalen Mannigfaltigkeiten},
  author={Stiefel, Eduard},
  year={1935},
  school={ETH Zurich}
}

@article{austin2021program,
  title={Program Synthesis with Large Language Models},
  author={Austin, Jacob and Odena, Augustus and Nye, Maxwell and Bosma, Maarten and Michalewski, Henryk and Dohan, David and Jiang, Ellen and Cai, Carrie and Terry, Michael and Le, Quoc and others},
  journal={arXiv preprint arXiv:2108.07732},
  year={2021}
}

@book{krzysztof2000generative,
  title={Generative Programming: Methods, Tools and Applications},
  author={Krzysztof, Czarnecki and Eisenecker, Ulrich W},
  year={2000},
  publisher={Addison-Wesley}
}

@article{bengio2003neural,
  title={A neural probabilistic language model},
  author={Bengio, Yoshua and Ducharme, R{\'e}jean and Vincent, Pascal and Jauvin, Christian},
  journal={Journal of machine learning research},
  volume={3},
  number={Feb},
  pages={1137--1155},
  year={2003}
}

@article{alon2019code2vec,
  title={code2vec: Learning distributed representations of code},
  author={Alon, Uri and Zilberstein, Meital and Levy, Omer and Yahav, Eran},
  journal={Proceedings of the ACM on Programming Languages},
  volume={3},
  number={POPL},
  pages={1--29},
  year={2019},
  publisher={ACM New York, NY, USA}
}

@inproceedings{takerngsaksiri2025code,
  title={Code readability in the age of large language models: An industrial case study from atlassian},
  author={Takerngsaksiri, Wannita and Tantithamthavorn, Chakkrit and Fu, Micheal and Pasuksmit, Jirat and Chen, Kun and Wu, Ming},
  booktitle={2025 IEEE International Conference on Software Maintenance and Evolution (ICSME)},
  pages={732--742},
  year={2025},
  organization={IEEE}
}

@book{spinellis2006code,
  title={Code quality: the open source perspective},
  author={Spinellis, Diomidis},
  year={2006},
  publisher={Adobe Press}
}

@inproceedings{mannan2018towards,
  title={Towards understanding code readability and its impact on design quality},
  author={Mannan, Umme Ayda and Ahmed, Iftekhar and Sarma, Anita},
  booktitle={Proceedings of the 4th ACM SIGSOFT International Workshop on NLP for Software Engineering},
  pages={18--21},
  year={2018}
}

@inproceedings{fakhoury2019improving,
  title={Improving source code readability: Theory and practice},
  author={Fakhoury, Sarah and Roy, Devjeet and Hassan, Adnan and Arnaoudova, Vernera},
  booktitle={2019 IEEE/ACM 27th International Conference on Program Comprehension (ICPC)},
  pages={2--12},
  year={2019},
  organization={IEEE}
}

@article{zhang2025style2code,
  title={Style2Code: A Style-Controllable Code Generation Framework with Dual-Modal Contrastive Representation Learning},
  author={Zhang, Dutao and Arias, Nicolas Rafael Arroyo and He, YuLong and Kovalchuk, Sergey},
  journal={arXiv preprint arXiv:2505.19442},
  year={2025}
}

\end{document}